\documentclass{aa}
\usepackage{graphicx}
\usepackage{txfonts}
\usepackage{aalongtable}
\usepackage{natbib}
\bibpunct{(}{)}{;}{a}{}{,} 
\begin{document}
   \title{Non-thermal radio emission from O-type 
          stars\thanks{Partly based on observations with \emph{XMM-Newton}, 
                       an ESA Science Mission with instruments and 
                       contributions directly funded by ESA Member
                       States and the USA (NASA).}}
   \subtitle{IV. Cyg OB2 No. 8A}

   \author{R. Blomme \inst{1}
           \and M. De Becker \inst{2,3}
           \and D. Volpi \inst{1}
           \and G. Rauw \inst{2}
          }


   \institute{Royal Observatory of Belgium,
              Ringlaan 3, B-1180 Brussel, Belgium \\
              \email{Ronny.Blomme@oma.be}
            \and
              Institut d'Astrophysique, Universit\'e de Li\`ege,
              All\'ee du 6 Ao\^ut, 17, B\^at B5c,
              B-4000 Li\`ege (Sart-Tilman), Belgium
            \and
              Observatoire de Haute-Provence,
              F-04870 Saint-Michel l'Observatoire,
              France
            }

   \date{Received date; accepted date}

   \abstract
   {Several early-type
    colliding-wind binaries are known to emit synchrotron radiation due to
    relativistic electrons, which are most probably
    accelerated by the Fermi mechanism.
    By studying such systems we can learn more about this mechanism,
    which is also relevant in other astrophysical contexts. 
    Colliding-wind binaries are furthermore
    important for binary frequency determination in clusters and
    for understanding clumping and porosity in stellar winds.
   }
   {We study the non-thermal radio emission of the binary Cyg~OB2 No.~8A,
    to see if it is variable and if that variability is locked to the
    orbital phase. We investigate if the synchrotron emission
    generated in the colliding-wind region of this binary can explain the
    observations and we verify that our proposed model is compatible
    with the X-ray data.}
   {We use both new and archive radio data from the Very Large Array (VLA)
    to construct a light curve as a function of orbital phase. We also
    present new X-ray data that allow us to improve the X-ray light curve.
    We develop a numerical model for the
    colliding-wind region and the synchrotron emission it generates.
    The model also includes free-free absorption and emission due to the
    stellar winds of both stars. In this way we construct artificial 
    radio light curves and compare them with the observed one.}
   {The observed
    radio fluxes show phase-locked variability. Our model can explain
    this variability because the synchrotron emitting region 
    is not completely hidden by the free-free absorption.
    In order to obtain a better agreement for 
    the phases of minimum and maximum flux we need to use
    stellar wind parameters for the binary components
    which are somewhat different from typical
    values for single stars. We verify that the change in stellar
    parameters does not influence the interpretation of the
    X-ray light curve. Our model has
    trouble explaining the observed radio spectral index. This could indicate
    the presence of clumping or porosity
    in the stellar wind, which 
    -- through its influence on both the Razin effect and the
    free-free absorption -- 
    can considerably influence the spectral index. 
    Non-thermal radio 
    emitters could therefore open a valuable pathway to investigate the 
    difficult issue of clumping in stellar winds.
    }
   {}


   \maketitle

\section{Introduction}

All hot stars emit radio radiation due to thermal free-free emission by the 
ionized material in their stellar wind. A number of those stars
also show
\emph{non-thermal} radio emission that dominates the thermal component.
This non-thermal emission is believed to be due to electrons that
are Fermi-accelerated around shocks \citep{1978MNRAS.182..147B}. 
As these relativistic electrons spiral around in the magnetic 
field, they
emit synchrotron radiation, which we detect as non-thermal radio 
emission \citep{1989ApJ...340..518B}.

Two hypotheses have been proposed
for the origin of the shocks responsible for the non-thermal emission. 
These shocks
could be due to the collision of the two winds in a binary 
system \citep{1993ApJ...402..271E, 2003A&A...409..217D, 2006A&A...446.1001P, 2006ApJ...644.1118R}, or
they could be formed by the radiative instability mechanism in the
wind of a single star 
\citep{1984ApJ...284..337O, 1994Ap&SS.221..259C}.
The binary hypothesis is supported by the non-thermal Wolf-Rayet stars
\citep{2000MNRAS.319.1005D} and by a number
of known O+O binaries showing non-thermal radio emission
\citep[e.g. \object{Cyg OB2 No. 5};][]{1997ApJ...488L.153C}.
The single-star hypothesis, on the other hand, is supported
by a number of seemingly single O stars showing non-thermal radio emission.

In recent years, however, doubt has been cast on the single-star hypothesis.
In previous papers of this series, we have shown that
the radio emission for some of these stars shows periodic variability, 
suggesting binarity 
\citep[\object{HD 168\,112} and \object{Cyg OB2 No. 9};][]
      {2005A&A...436.1033B, 2008A&A...483..585V}.
Optical spectroscopy has shown that a number of these ``single" stars 
are actually binaries 
\citep[e.g. \object{Cyg OB2 No.~8A} and No. 9;][]
      {2004A&A...424L..39D, 2008A&A...483..543N}.
Furthermore, theoretical work by
\citet{2006A&A...452.1011V} showed that non-thermal
radio radiation from a single star would be unlikely, as the
synchrotron emission would be largely absorbed by 
the stellar wind material.

The study of colliding-wind systems is important for a number
of reasons. First, because we can learn
more about the Fermi acceleration mechanism, which
is relevant to a wide
range of astrophysical objects such as interplanetary shocks and
supernova remnants \citep[e.g.][]{1991SSRv...58..259J}.
The physical conditions
in a colliding-wind binary are quite different from those in other objects.
The different range of density, magnetic field
and radiation environment in these binaries provides substantial new
information on the acceleration mechanism.
Models, constrained
by radio data, also provide predictions for the expected gamma-ray flux
from these systems. These numbers are important for observations
of colliding-wind binaries by current and future high-energy instruments,
both
ground-based and space-based ones (e.g. INTEGRAL, IXO, HESS, VERITAS).

Furthermore, the non-thermal radio emitters are useful
for improving the determination of the binary frequency in clusters.
The colliding-wind binaries allow us to find the
longer-period systems, which are easy to detect at radio wavelengths
but much harder to find
with standard spectroscopic observing techniques. 

Finally, colliding-wind systems are also highly relevant for
mass-loss rate determinations in single stars. 
This has become a major problem in 
massive star research in recent years due to the fact that 
winds are clumped and porous. Taking into account clumping
decreases the mass loss rates by a factor $3-10$ compared
to what was previously thought
\citep[e.g.][]{2008A&ARv..16..209P}.
By determining how much of the synchrotron radiation emitted at the
colliding-wind region is absorbed by the stellar wind material,
we can estimate the amount of porosity in the wind.

In this paper we study the non-thermal radio emission of the 
colliding-wind binary Cyg~OB2 No.~8A (RA = 20h33m15{\fs}0789, 
Dec = +41{\degr}18'50{\farcs}494, J2000).
We also present additional X-ray observations that improve the
X-ray light curve presented by \citet{2006MNRAS.371.1280D}.
Cyg~OB2 No.~8A is undoubtedly the O + O
non-thermal radio emitter with the best constrained stellar, wind and
orbital parameters so far \citep{2007A&ARv..14..171D}.
It has been known as a non-thermal radio emitter since the
first major survey of O star radio emission by
\citet{1989ApJ...340..518B}. It was recognized as such by its clearly
negative spectral index.
This system was only recently discovered to be a binary
by \citet{2004A&A...424L..39D}. It consists of
an O6If primary and an O5.5III(f) secondary. It 
has a period of
21.908 $\pm$ 0.040 d and an eccentricity of 0.24 $\pm$ 0.04.
All binary parameters were determined by \citet{2006MNRAS.371.1280D} and
are listed in Table~\ref{table parameters}.

\begin{table}
\caption{Parameters of Cyg~OB2 No.~8A used in this work.}
\label{table parameters}
\centering
\begin{tabular}{llllll}
\hline\hline
Parameter & Primary & Secondary \\
\hline
$T_{\rm eff}$ (K)                & 36800 & 39200 \\
$R_*$ ($R_{\sun}$)                 & 20.0  & 14.8 \\
$M_*$ ($M_{\sun}$)                 & 44.1  & 37.4 \\
log $L_{\rm bol}/L_{\sun}$         & 5.82  & 5.67 \\
$\dot{M}$ ($M_{\sun} {\rm yr}^{-1}$) & $4.8 \times 10^{-6}$ & $3.0 \times 10^{-6}$ \\
$\varv_\infty$ (km s$^{-1}$)           & 1873  & 2107  \\
semi-major axis $a$ ($R_{\sun}$) & 65.7 & 76.0 \\

period $P$ (days)  & \multicolumn{2}{c}{$21.908 \pm 0.040$} \\
eccentricity $e$   & \multicolumn{2}{c}{$0.24 \pm 0.04$} \\
inclination $i$    & \multicolumn{2}{c}{$32\degr \pm 5\degr$} \\
distance $D$ (kpc) & \multicolumn{2}{c}{1.8} \\
\hline
\end{tabular}
\tablefoot{These parameters were derived by \citet{2006MNRAS.371.1280D}.}
\end{table}

\begin{table*}
\begin{center}
\caption{Two new X-ray observations of Cyg~OB2 No.~8A. \label{table xraydata}}
\begin{tabular}{ccclcccccccc}
\hline\hline
Phase & Rev.    & Obs. ID & Date & JD & \multicolumn{3}{c}{Exp. Time} &  & \multicolumn{3}{c}{Count Rates} \\
      &         &         &      &    & \multicolumn{3}{c}{(s)} &  & \multicolumn{3}{c}{(counts\,s$^{-1}$)} \\
\cline{6-8}\cline{10-12}
 &    &  &  &  & MOS1 & MOS2 & pn & & MOS1 & MOS2 & pn \\
\hline
0.155 & 1353 & 0505110301 & 2007 Apr 29 & 4220.417 & 26879 & 27149 & 20581 & & 0.655$\pm$0.006 & 0.640$\pm$0.005 & 1.892$\pm$0.012 \\
0.326 & 1355 & 0505110401 & 2007 May 03 & 4224.167 & 27065 & 29155 & 22116 & & 0.734$\pm$0.006 & 0.730$\pm$0.006 & 2.114$\pm$0.011 \\
\hline
\end{tabular}
\tablefoot{These X-ray observations were performed in 2007 with
\emph{XMM-Newton}. The first five columns give respectively
the orbital phase, the revolution
number, the observation ID, the observation date, and the Julian date
($-$2\,450\,000) at mid-exposure. The other columns give respectively
the effective exposure time, and the count rates
for the three EPIC instruments. }
\end{center}
\end{table*}

\begin{table*}
\caption{Parameters for the two new EPIC spectra of Cyg~OB2 No.~8A.
\label{table xrayfit}}
\begin{center}
\begin{tabular}{cccccccccc}
\hline
Instr.  & Log\,$N_\mathrm{w}$ & $kT_1$ & Norm$_1$ & $kT_2$ & Norm$_2$ & $kT_3$ & Norm$_3$ & $\chi^2_\nu$ (d.o.f.) & Obs.Flux \\
	&  & (keV) & (10$^{-2}$) & (keV) & (10$^{-3}$) & (keV) & (10$^{-3}$) &  & (erg\,cm$^{-2}$\,s$^{-1}$) \\
\hline
\multicolumn{9}{l}{{\it Observation 5}}\\
\hline
MOS1	& 21.79$_{21.68}^{21.87}$ & 0.26$_{0.23}^{0.30}$ & 3.43$_{1.39}^{6.76}$ & 0.78$_{0.72}^{0.87}$ & 7.81$_{6.43}^{9.47}$ & 1.73$_{1.60}^{1.90}$ & 6.33$_{5.31}^{7.24}$ & 1.28 (230) & 5.28\,$\times$\,10$^{-12}$ \\ 
pn	& 21.86$_{21.80}^{21.91}$ & 0.25$_{0.23}^{0.29}$ & 6.06$_{3.43}^{10.10}$ & 0.84$_{0.79}^{0.89}$ & 10.96$_{9.48}^{12.44}$ & 1.86$_{1.71}^{2.06}$ & 4.96$_{3.91}^{5.94}$ & 0.93 (436) & 5.69\,$\times$\,10$^{-12}$ \\
EPIC	& 21.81$_{21.77}^{21.86}$ & 0.27$_{0.25}^{0.29}$ & 3.72$_{2.48}^{5.78}$ & 0.81$_{0.77}^{0.86}$ & 8.85$_{7.93}^{9.90}$ & 1.77$_{1.68}^{1.88}$ & 5.80$_{5.09}^{6.44}$ & 1.36 (691) & 5.49\,$\times$\,10$^{-12}$ \\
\hline
\multicolumn{9}{l}{{\it Observation 6}}\\
\hline
MOS1	& 21.80$_{21.72}^{21.84}$ & 0.22$_{0.20}^{0.24}$ & 8.14$_{4.66}^{12.20}$ & 0.89$_{0.82}^{0.99}$ & 7.02$_{5.79}^{8.64}$ & 1.60$_{1.51}^{1.75}$ & 7.96$_{6.12}^{9.20}$ & 1.78 (329) & 6.00\,$\times$\,10$^{-12}$ \\ 
pn	& 21.73$_{21.68}^{21.77}$ & 0.26$_{0.24}^{0.28}$ & 3.83$_{2.70}^{5.50}$ & 0.79$_{0.76}^{0.82}$ & 9.70$_{8.86}^{1.79}$ & 10.56$_{1.67}^{1.79}$ & 7.75$_{7.24}^{8.23}$ & 1.11 (843) & 6.72\,$\times$\,10$^{-12}$ \\
EPIC	& 21.70$_{21.67}^{21.72}$ & 0.26$_{0.25}^{0.28}$ & 3.17$_{2.53}^{3.84}$ & 0.81$_{0.78}^{0.84}$ & 7.60$_{7.06}^{8.14}$ & 1.68$_{1.63}^{1.72}$ & 8.07$_{7.66}^{8.45}$ & 1.96 (1274) & 6.39\,$\times$\,10$^{-12}$ \\
\hline
\end{tabular}
\tablefoot{The parameters are derived from a
{\tt wabs$_\mathrm{ISM}$*wind*(mekal$_1$+mekal$_2$+mekal$_3$)} model.
Results are given for the MOS1, pn, and combined MOS1+pn (`EPIC') instruments.
The first absorption component
({\tt wabs$_\mathrm{ISM}$}) is frozen at the ISM value:
0.94\,$\times$\,10$^{22}$ cm$^{-2}$. The second absorption column,
quoted as $N_\mathrm{w}$ (in cm$^{-2}$), stands for the absorption by the
ionized wind material. The normalization parameter (Norm) of the
{\tt mekal} components is defined as
$(10^{-14}/(4\,\pi\,D^2))\int{n_\mathrm{e}\,n_\mathrm{H}\,dV}$, where $D$,
$n_\mathrm{e}$ and $n_\mathrm{H}$ are respectively the distance to the
source (in cm), and the electron and hydrogen number densities (in cm$^{-3}$).
The indicated range in the parameter values represents the 90\,\% confidence
interval. The last column gives the observed flux between 0.5 and 10.0 keV.
}
\end{center}
\end{table*}

\citet{2006MNRAS.371.1280D} studied the X-ray emission of
this binary.  They found an essentially thermal spectrum, but with an
overluminosity of a factor $19-28$ compared to the canonical $L_X/L_{\rm bol}$
ratio for non-interacting O-type stars. This overluminosity is due to additional
X-ray emission by material heated
in the colliding-wind region. The X-ray light curve
shows variability of $\sim$~20\,\%. Although the curve is not well sampled,
it suggests that the variability is phase-locked.

In Sect.~\ref{sect radio data} we present the radio continuum observations 
of Cyg~OB2 No.~8A and in Sect.~\ref{sect X-ray data} the new X-ray data.
The model for the synchrotron radio emission
in a colliding-wind binary is described in Sect.~\ref{sect modelling}. 
In Sect.~\ref{sect results} we apply this model both for a standard set
of star and wind parameters and an alternative set chosen to obtain
better agreement with the data. Conclusions are given in 
Sect.~\ref{sect conclusions}.

\section{Radio data}
\label{sect radio data}

We obtained a number of 6~cm continuum observations of
Cyg~OB2 No.~8A with the 
NRAO\footnote{The National Radio Astronomy
             Observatory is a facility of the National Science Foundation
             operated under cooperative agreement by Associated Universities,
             Inc.}
Very Large Array (VLA) during 2005 February 04 to March 12, covering about 1.6 
orbital periods. We also retrieved a considerable number of observations
from the VLA archive (also covering other wavelengths), 
which we used in the present analysis. To avoid
introducing systematic effects, we re-reduced the archive data in the
same way as our own data (see Appendix~\ref{appendix data reduction}).
The resulting fluxes and their error bars are listed in 
Table~\ref{table radio data}.

\begin{figure*}
\sidecaption
\includegraphics[width=12cm]{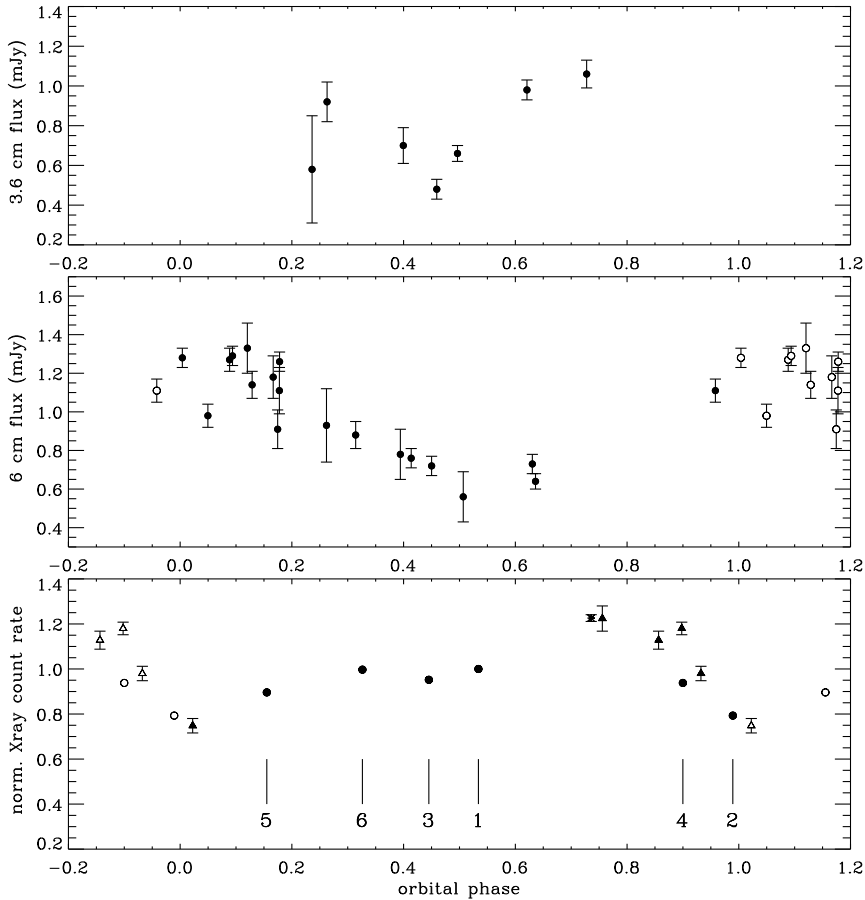}
\caption{Radio and X-ray observations of
Cyg~OB2 No.~8A. 
The upper and middle panel show the 
3.6~cm and 6~cm radio fluxes.
We plot only those fluxes that were measured on-centre.
All fluxes are plotted
as a function of orbital phase in the 21.908-d period, with
phase 0.0 corresponding to periastron.
The lower panel shows 
the X-ray fluxes, with count rates that have been normalized on 
\emph{XMM-Newton} observation no. 1 (see text). 
Triangles indicate \emph{ROSAT}-PSPC observations, the cross the
\emph{ASCA}-SIS1 observation, and circles the \emph{XMM-Newton}-EPIC data
(these are also labelled by their observation number). 
Error bars are shown, except for the \emph{XMM-Newton} data, 
as these are smaller than the symbol used.
To better show the behaviour of the X-ray and radio
light curves, we extend the phase
range by 0.2 on either side;
open symbols indicate duplications in that extended range.
}
\label{figure fluxes}
\end{figure*}

In Fig.~\ref{figure fluxes}, we plot a subset of these fluxes.
We limit the figure to the 3.6 and 6~cm data as the other wavelengths
have only a few observations or detections.
We do not plot fluxes which were measured on images where Cyg~OB2 No.~8A
was offset from the centre. Though these observations follow the
same general trend as the ``on-centre" observations, 
they usually have large error bars and 
would therefore considerably confuse the figure.

Figure~\ref{figure fluxes} shows clear variability in the fluxes which is
phase-locked with the orbital period. This is especially
obvious at the 6~cm wavelength which has the most data.
Maximum flux is at phase $\sim$~0.1 (where the primary is in front). 
The phase of minimum flux is less
well determined due to a gap in the phase coverage, but is at
$\sim$~0.6 (close to where the secondary is in front).
To check that the variability is not an instrumental effect,
we also measured two other targets on the images: 
\object{Cyg~OB2 No.~63} and No.~9. Although these are
also variable 
\citep[No.~9, see][]{2008A&A...483..585V},
their variability is not correlated with the No.~8A variability.
We can therefore exclude instrumental effects.

Only a few simultaneous observations at more than one wavelength are
available to study the spectral index. We find that 
before phase 0.5, the 3.6 and 6~cm fluxes are roughly equal,
corresponding to a spectral index of about 0.0
(within the error bars).
After phase 0.5, the fluxes behave differently: 
the 3.6~cm flux reaches
higher values much more quickly than the 6~cm one, which translates
into a spectral index of $\sim$\,$+0.7$. 
Interestingly, this is close to the orbital phase
where the only significantly negative
spectral index ($\sim$\,$-0.7$) is found, occurring between
the 6 and 20 cm observations
of programme AC116 (1984-12-21, phase = 0.51).

We next try to determine the period from the 6~cm data used in 
Fig.~\ref{figure fluxes}. To do so, we use the
string-length method of
\citet{1983MNRAS.203..917D}. The best period we find is
23.75 d, which is quite different from the 21.908-d value
derived from optical spectroscopy \citep{2004A&A...424L..39D}.
However, the minimum in the string-length is not very significant.
Furthermore, if we replot the \citeauthor{2004A&A...424L..39D} 
spectroscopic velocities 
on the 23.75-d period, we find that the points are
scattered all over the diagram, indicating that the proposed radio
period is
clearly wrong. We therefore use the 21.908-d period in
the further analysis.

\section{X-ray data}
\label{sect X-ray data}

Cyg~OB2 No.~8A has been observed in X-rays two times with the \emph{XMM-Newton} 
satellite since the publication of the \cite{2006MNRAS.371.1280D} paper 
(Obs. ID 050511, PI: G. Rauw). These observations were scheduled in order 
to fill the gap in phase coverage of previous X-ray observations, 
and occurred during revolutions 1353 and 1355 (respectively at phases 
0.155 and 0.326, see Table~\ref{table xraydata} for details). 
We adopted the same procedure as in \citeauthor{ 2006MNRAS.371.1280D}
for the data processing using the Science Analysis Software (v.6.0.0), 
and we refer to this study for details. 
We note that we rejected short time intervals of high background due to 
soft proton flares.
We fitted the new EPIC spectra 
with the same three-temperature model as used by 
\citet{2006MNRAS.371.1280D}
for previous {\it XMM-Newton} observations. The best fit parameters are 
given in Table\,\ref{table xrayfit};
they are consistent with 
those derived for the four previous observations, even though some 
variability is detected for some of them. 

\begin{figure}
\begin{center}
\includegraphics[width=8cm]{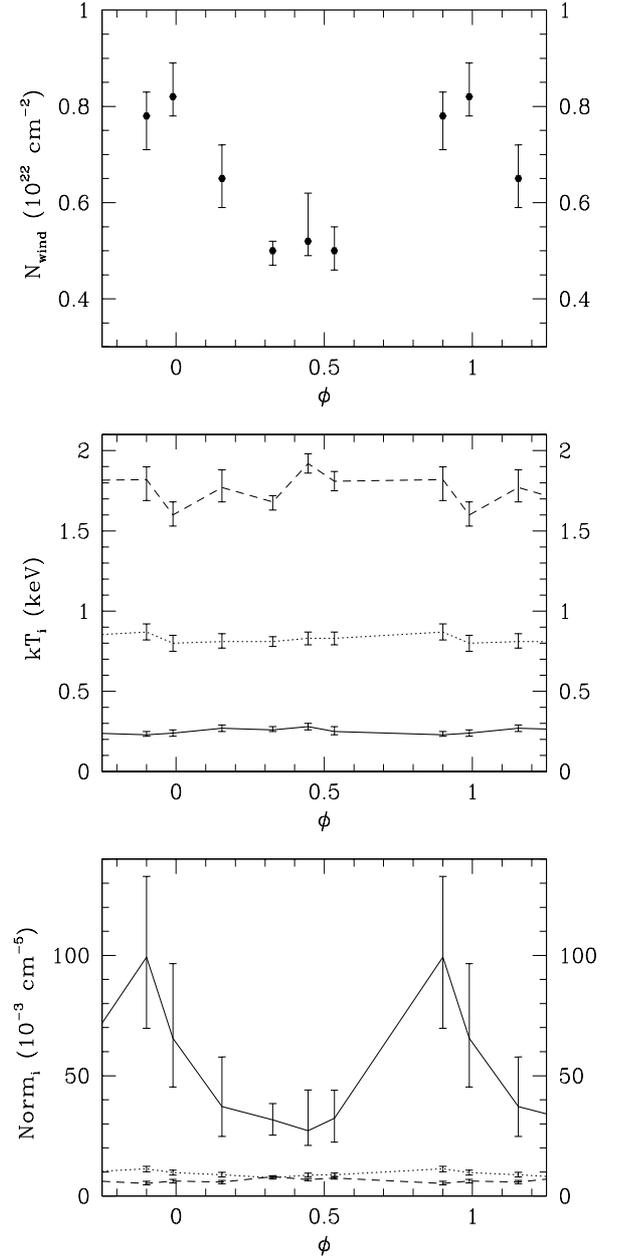}
\caption{
Variation of the best fit parameters as a function of orbital phase
($\phi$), for all six EPIC X-ray spectra. 
Upper panel: local absorption components expressed in 
10$^{22}$\,cm$^{-2}$. Middle panel: plasma temperature expressed in 
keV. Lower panel: normalization parameters expressed in units of 
10$^{-3}$\,cm$^{-5}$. In the middle and lower panels, the solid, dotted and 
dashed lines stand, respectively, for the soft, the medium and the hard 
emission components of the composite model.
\label{fig nwind}}
\end{center}
\end{figure}

In Fig.~\ref{fig nwind}, we have plotted the parameters obtained from the 
best fit of EPIC data (MOS1 and pn) of the six {\it XMM-Newton} observations 
(see Table~4 in \citet{2006MNRAS.371.1280D},
and Table~\ref{table xrayfit} in this paper) versus the orbital phase.  
The local absorption column density undergoes significant variability as a 
function of the orbital phase, with a maximum occurring close to periastron 
passage (phase 0.0). At that time 
the wind interaction region is behind the primary wind, which is
likely to be responsible for a substantial absorption of X-rays. The plasma 
temperatures show only weak, or even no phase-locked variability. 
Only slight variations are suggested in the case of the hard emission 
component that might be related to the small changes in the pre-shock 
velocities as the separation between the stars changes along the eccentric 
orbit. The lower and higher temperature are indeed found, respectively, 
close to periastron and apastron for the hard component, but this trend should 
be considered with caution in view of the scatter and of the error bars of 
the measurements. Finally, the normalization parameters 
(which are related to the emission measure) undergo significant 
phase-locked variations, mostly if one considers the soft emission component 
that dominates the X-ray spectrum. The maximum coincides remarkably with 
orbital phases close to periastron, in agreement with the idea that the 
maximum X-ray emission occurs when the emission measure is expected to be 
maximum.

We proceeded following the same approach as in \cite{2006MNRAS.371.1280D}
to build a light curve including \emph{ROSAT}-PSPC, \emph{ASCA}-SIS and 
\emph{XMM-Newton}-EPIC data (see their Fig.~6). 
In order to consistently compare the count rates obtained with different
instruments, we used the first \emph{XMM-Newton} observation as a reference
(obs.~1 in \citeauthor{2006MNRAS.371.1280D}, 
obtained at orbital phase 0.534). We used the 3-temperature 
model with the parameters obtained for the simultaneous fit of
EPIC-MOS1 and
EPIC-pn data of that observation, and we convolved it with the respective
response matrices of \emph{ROSAT}-PSPC and \emph{ASCA}-SIS1 
to obtain faked spectra. We
then normalized the count rates of the \emph{ROSAT} and \emph{ASCA} 
observations using
the count rate of the reference fake spectrum in each case, thereby deriving
the normalized count rates plotted in 
Fig.~\ref{figure fluxes}.
It is interesting to note that the X-ray emission level at orbital phases 
0.155 and 0.326 (respectively no.~5 and no.~6 in the \emph{XMM-Newton} series) 
is not lower than at
other orbital phases. This suggests that the X-ray minimum occurs very close to periastron.

An important result of 
Fig.~\ref{figure fluxes} is that the X-ray and radio light curves are
nearly anti-correlated. X-ray flux minimum occurs close to periastron,
when the radio 6~cm flux is approaching maximum. Minimum radio flux
is near phase 0.6, where the X-ray flux has increased well above
its minimum value (though it has not yet reached maximum). The anti-correlation
will present a challenge to modelling this binary system. Although we
do not model the X-ray observations in the present paper, we discuss this
point further in Sect.~\ref{section X-ray emission}.

\section{Modelling the radio fluxes}
\label{sect modelling}

\citet{2005mshe.work...45B}
made a very simplified model for the radio emission of the Cyg~OB2 No.~8A binary
system. This model consists of a point-source synchrotron emitting region
orbiting a single star inside a stellar wind that includes free-free
absorption. The model shows that all synchrotron emission would be
absorbed by the stellar wind material and that therefore 
the radio fluxes should not show phase-locked variability.
Only by introducing a considerable
amount of porosity in the stellar wind material
does the non-thermal radio emission become detectable
(due to the reduction in free-free absorption).

Here we present a more sophisticated model of Cyg~OB2 No.~8A
that takes into account the geometric extent of the colliding-wind region,
as well as the acceleration, advection and cooling of the
relativistic electrons.
In this section we provide an outline of the model; for 
details we refer to
Appendix~\ref{appendix modelling}.

The present model does not solve the hydrodynamical equations, but
instead uses the \citet{2004ApJ...611..434A} equations
to define the position of the
contact discontinuity that separates the two stellar winds.
Estimates by \citet[][their Table 11]{2006MNRAS.371.1280D} show the ratio of
cooling time to the flow time to be around~1. At least during part of the
orbit, the shocks can therefore be expected to be radiative, resulting
in considerable variability and structure. Such complications
are neglected in our model.

Next we assume that the contact discontinuity and the shocks on either side 
of it are sufficiently close to one another that we can consider
them to be in the same geometric place.
We then generate new relativistic electrons at the 
shocks and follow their time evolution as they advect and cool,
till they no longer emit significant synchrotron radiation at the wavelength
we are considering.

At the shock, the new electrons have a momentum distribution
that is a power law modified for the effect of inverse Compton cooling.
The number of relativistic electrons is determined by $\zeta$, the
fraction of the energy that is transferred from the shock to the
relativistic electrons. We assume the typical value of
$\zeta=0.05$
\citep{1987PhR...154....1B, 1993ApJ...402..271E}.
The time evolution of the electrons 
is followed as they advect outward
along the contact discontinuity and as they cool down due to inverse
Compton and adiabatic cooling.

At each point along 
the tracks followed by the electrons,
the synchrotron emissivity
$j_\nu (r)$ can then be calculated. We include the Razin effect, which
takes into account the effect of the non-relativistic plasma on
the synchrotron emission.
This emissivity is then mapped from the
two-dimensional orbital plane on which it was defined on to a 
three-dimensional volume by rotating it along
the line connecting both components of the binary. 
Note that this assumption of cylindrical symmetry neglects
any effects due to orbital motion.
We refer to \cite{2007ApJ...662..582L}, \cite{2008MNRAS.388.1047P}
and \cite{2009MNRAS.396.1743P} for studies of the orbital motion effects
in colliding-wind systems in general. Specifically for 
the relatively short-period system Cyg~OB2 No.~8A,
not taking into account orbital motion is a
major assumption, and we discuss its effect in 
Sect.~\ref{section improved model}.

The model also includes the thermal free-free opacity and 
emissivity due to the ionized wind material.
The winds are assumed to be smooth, i.e. no clumping or porosity effects
are included. The model
does not
include the contribution due to the increased density and
temperature in the colliding-wind region 
\citep[contrary to][]{2010MNRAS.403.1633P}
as these hydrodynamical quantities are
not calculated in our model.
Radiative transfer is
then solved in the three-dimensional simulation volume
using Adam's finite volume method
\citep{1990A&A...240..541A}.

\section{Results}
\label{sect results}

The intention of our modelling is to see if we can reproduce qualitatively
the main features of the observed Cyg~OB2 No.~8A radio light curve
(see Sect.~\ref{sect radio data}). 
These are: 
(1) the fact that there is variability, 
(2) maximum flux is around phase 0.1, minimum flux around phase 0.6,
and (3) the 3.6~cm and 6.0 cm fluxes are approximately equal,
most of the time (i.e. spectral index $\approx 0$).
We do not make a formal best fit as we 
consider the absolute values of the fluxes to be less important,
since these are sensitive to many of the assumptions in the model
(e.g. magnetic field, $\zeta$).

\begin{figure}
\resizebox{\hsize}{!}{\includegraphics{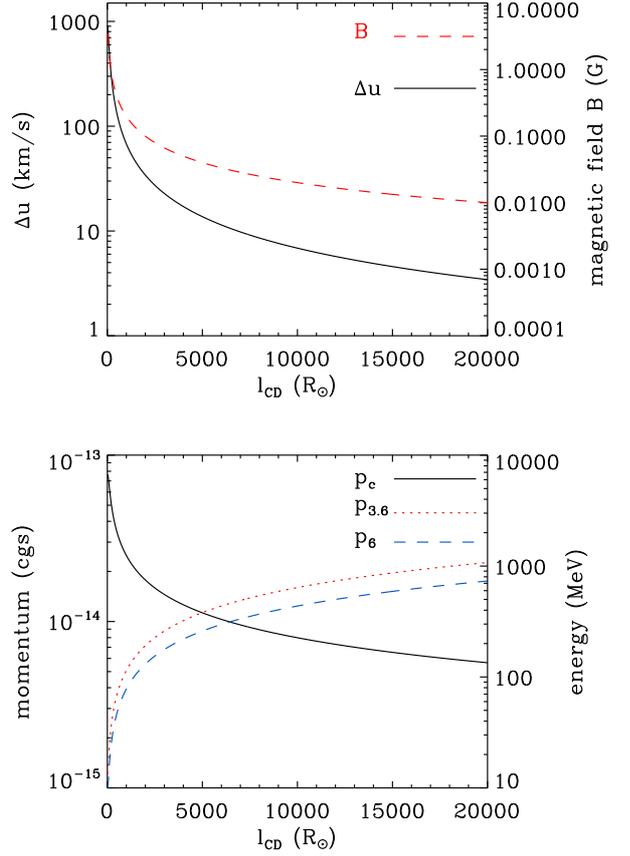}}
\caption{
Upper panel: The jump in velocity at the shock ($\Delta u$, solid line)
as a function of distance along the contact discontinuity
($l_{\rm CD}$). Also
plotted is the magnetic field ($B$, dotted line). Lower panel:
The maximum momentum ($p_{\rm c}$) due to inverse Compton cooling
of the electrons accelerated at the shock,
as a function of distance (solid line). Also plotted are the
characteristic momenta for emission at 3.6~cm (dotted line) and
6~cm (dashed line).
All data presented here are for the wind of the primary component of the 
binary, at periastron.
}
\label{fig momentum}
\end{figure}

\subsection{Standard model}
\label{section standard model}

\subsubsection{Model details}

The standard model uses the parameters listed in 
Table~\ref{table parameters}. 
We recall that \citet{2006MNRAS.371.1280D} derived these from stellar
parameters typical for single stars with the same spectral type as the
binary components of Cyg~OB2 No.~8A.
We assume a surface magnetic field for
each star of $B=100$~G, which is compatible with the fact that 
the magnetic field of most O-type stars is below detectable levels
\citep[e.g.][]{2008A&A...483..857S}. 
The magnetic field at the shocks is directly related to this
surface magnetic field (see Eq~\ref{eq magnetic field simple}
and Fig.~\ref{fig momentum}, upper panel); 
this neglects any amplification
of the field \citep[such as described by, e.g.][]{2004MNRAS.353..550B}
or any reduction of it by magnetic reconnection.
As the true rotational velocity is not known,
we make the assumption that $\varv_{\rm rot}/\varv_\infty=0.1$.

\begin{figure*}
\centering
\includegraphics[width=17cm]{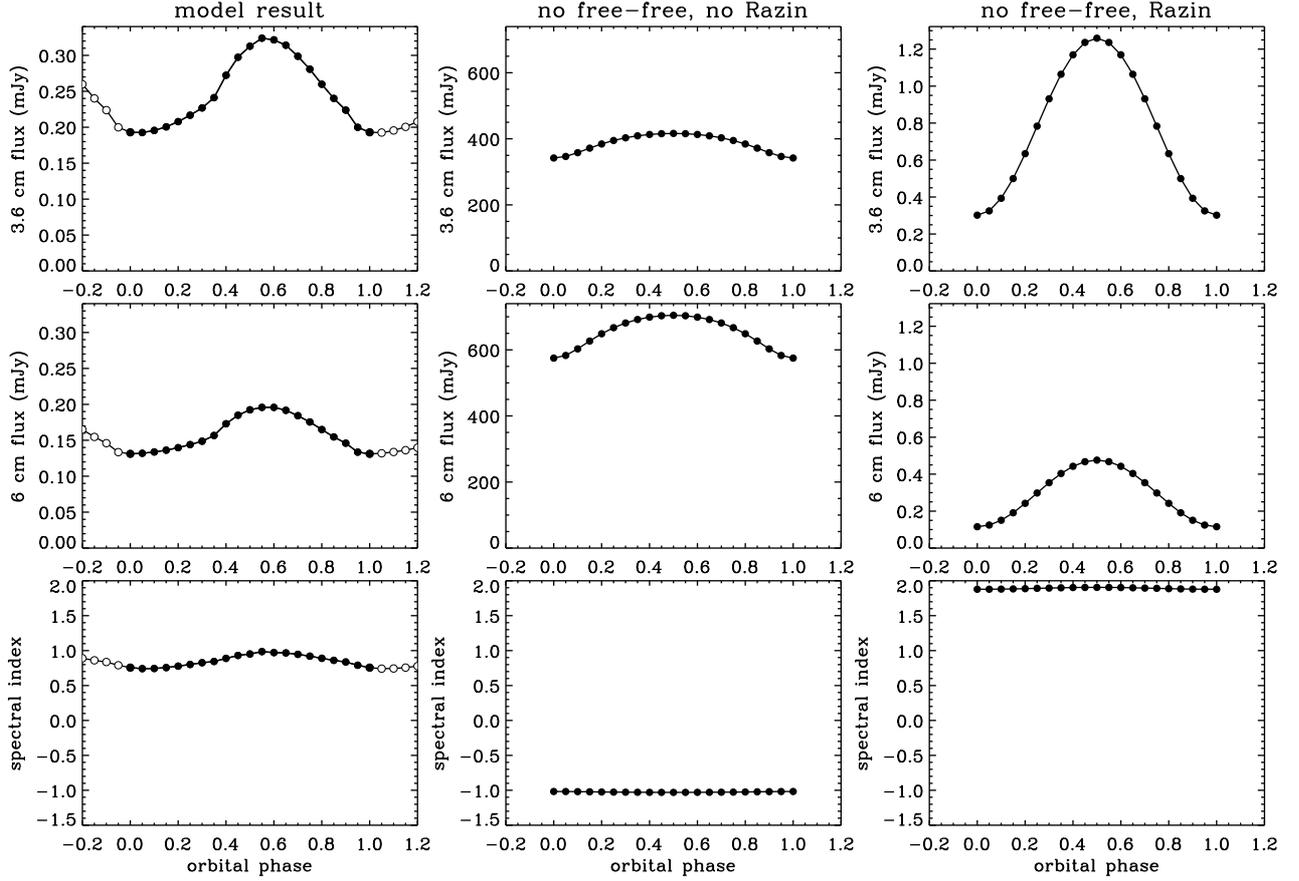}
\caption{Results for the standard model. The left column shows
the 3.6 cm (upper panel) and 6 cm fluxes (middle panel), 
as well as the spectral index
between those wavelengths (lower panel). The middle column shows the
same model, but we switched off the Razin effect and the
free-free absorption and emission. The right column shows the intrinsic
synchrotron emission, i.e. the Razin effect is included, but not the
free-free absorption and emission. 
As for Fig.~\ref{figure fluxes}, open symbols indicate duplications
in the extended phase range. 
Note the very different flux
scales for the various panels.}
\label{fig results}
\end{figure*}

A comparison between the $m \sin^3 i$ values derived from the 
binary orbit and typical masses of single stars corresponding to the 
observed spectral types indicates an inclination angle of 
32\degr~\citep{2006MNRAS.371.1280D}. For the free-free emission and absorption
we take the material to be once ionized and assign it a temperature
of 28\,500 K ($\approx 0.75$ of the effective temperature of the stars).  
The three-dimensional simulation volume is a cube centred on the primary
that is $40\,000\,R_{\sun}$ on each side and that consists of $256^3$ cells.
Some care must be taken to use the appropriate resolution
for a given size of the simulation box;
we checked that the results presented here are robust to changes in 
size and resolution.
The size of the box is large enough to contain most, but not all, 
of the emission. Doubling the size and using $512^3$ cells
shows that the flux loss is less than 10~\%, and that 
the shape of the radio light curve
remains the same.
With the resolution we use for the three-dimensional
simulation volume, the apex of the contact discontinuity is not 
well resolved. This is of no consequence for our purpose, however,
as the synchrotron emission coming from around the apex is
completely absorbed by the free-free absorption (see also 
Sect.~\ref{sect map into 3D}).

\subsubsection{Fluxes and spectral index}

Figure~\ref{fig results} (left column) shows the resulting
3.6 cm and 6 cm fluxes as well as the spectral index between 
these two wavelengths. The first thing to note is that the
model fluxes do show phase-locked variability, contrary to the 
simple point-source model \citep{2005mshe.work...45B}. 
This is due to the 
synchrotron emission extending much further out than the point-source
approximation suggests. In Fig.~\ref{fig cumulative flux} 
(left column)
we plot the contribution to the flux as a function of distance to the
primary. To determine each flux value on that figure we artificially set the
synchrotron emission to zero beyond a certain distance and re-run the 
radiative transfer model. 
The figure makes clear that the main flux contribution comes from distances of
$1000 - 3000\,R_{\sun}$.
This corresponds to approximately $10 - 30$ times the separation between
the two stars.
The flux contribution that can be seen at small distances
is due to free-free emission of the entire wind, 
which was not cut off in this procedure.
On the figure we also indicate the distances from the star where
optical depth 1 for the primary component
is reached. We calculate these using the 
\cite{1975MNRAS.170...41W} formalism and find them to be
$1130\,R_{\sun}$ at 3.6~cm
and $1620\,R_{\sun}$ at 6~cm. As expected, the flux we receive is formed
largely outside this radius.

Spatially resolved radio observations of colliding-wind regions typically
do not show such an extended non-thermal region 
\citep[e.g. WR 140;][]{2005ApJ...623..447D}. The present results are not
in contradiction with this. The resolved observations show the surface
brightness
(flux per area) while the unresolved radio data presented
here show the flux (i.e. integrated over a large area).
At large distances from the centre of the binary system, 
the surface brightness decreases
substantially, hence the relatively small extent of the colliding-wind
region in resolved 
observations 
-- which is nicely illustrated by the simulated data for WR~140
presented by \citet[][their Fig.~17]{2006MNRAS.372..801P}.
But, as the synchrotron emission is 
cylindrically symmetric around the axis connecting the two stars, 
the low surface brightness at larger distances 
gets weighted with a much larger surface area and therefore contributes
substantially to the flux.

While the phase-locked
variability of the flux is a positive result for the model, 
unfortunately we find that the phases of
minimum and maximum flux do not agree with the observations.
The model results are nearly in anti-phase to the observations:
theoretical maximum occurs at phase 0.55 (instead of the observed
0.1) and theoretical minimum at phase $0.0 - 0.05$ (instead of 0.6).
Furthermore, the spectral index is not correctly predicted. The theoretical
values of +0.75 to +1.0
do not agree with the observed values which are
mostly close to 0.0 (Sect.~\ref{sect radio data}).

\subsubsection{Without Razin effect, without free-free absorption}
\label{sect without Razin without free-free}

To understand why the phases and spectral index are incorrectly modelled, 
we start by simplifying the model.
When we leave out the free-free absorption and emission as well as the
Razin effect, we find the light curve shown in the middle column on 
Fig.~\ref{fig results}.
The flux values are high ($\sim$~700 mJy at 6~cm, $\sim$~400~mJy
at 3.6~cm). The spectral index is approximately $-1.0$, clearly
indicating the non-thermal nature of the 
spectrum\footnote{
The spectral index for a pure power-law momentum distribution $p^{-n}$
is $-(n-1)/2$, with
$n=(\chi+2)/(\chi-1)$. As we use $\chi=4$, one would expect a $-0.5$ spectral
index, but \citet[][their Fig.~4]{2006A&A...446.1001P} show that
the inclusion of inverse Compton cooling changes this index to a value
closer to $-1.0$.
}. 

\begin{figure*}
\includegraphics[width=17cm]{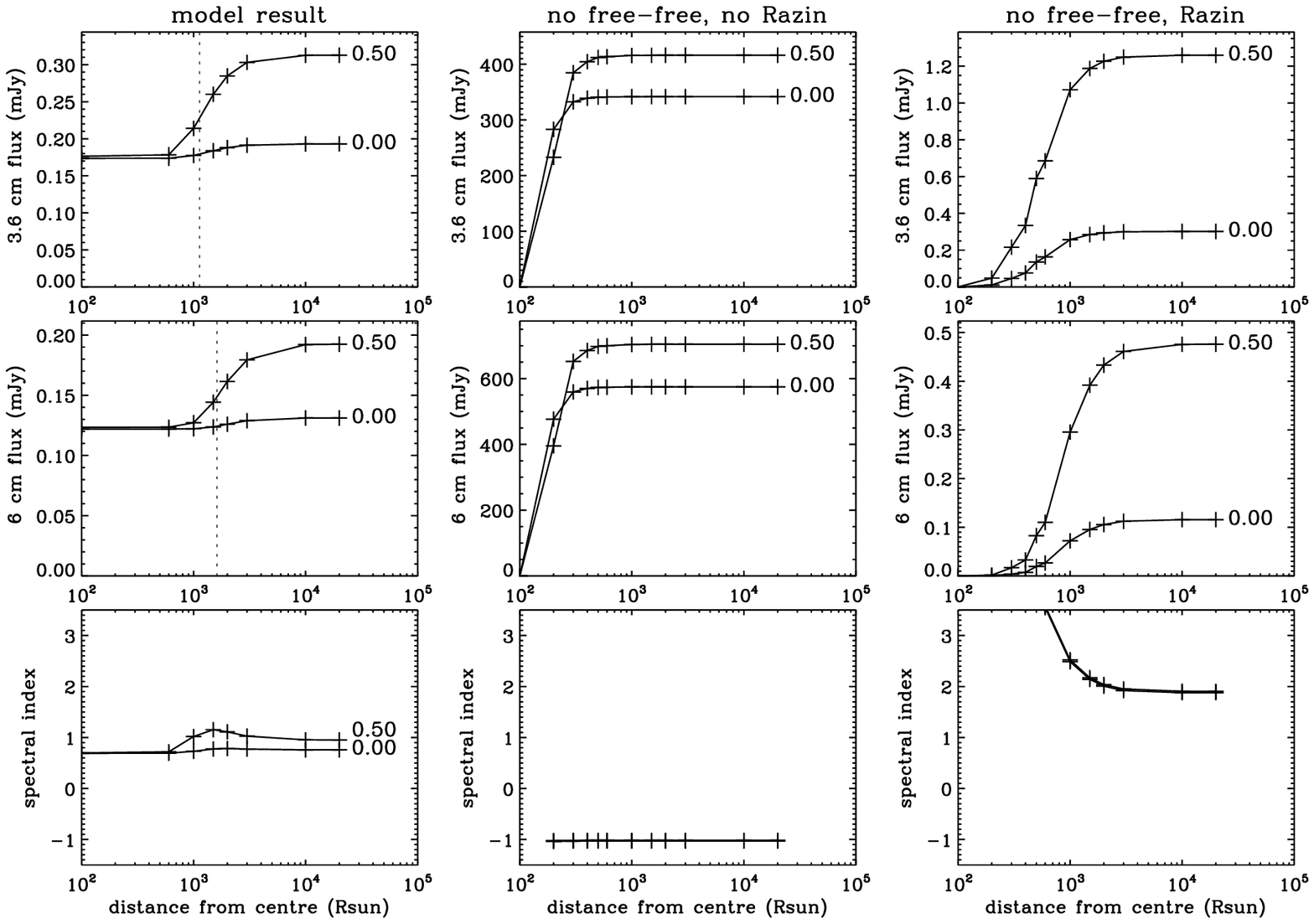}
\caption{Cumulative contribution to the flux from synchrotron emission
at various distances, for orbital phases 0.0 and 0.5. 
The left column shows the 3.6 cm (upper panel) and 6 cm fluxes (middle panel), 
as well as the spectral index between those wavelengths (lower panel). 
The dotted lines show where optical depth 1 is 
reached due to free-free absorption (for the wind of the primary component). 
The middle column shows the same model, but we switched off the Razin effect 
and the free-free absorption and emission. The right column shows the intrinsic
synchrotron emission, which includes the Razin effect, but not the
free-free absorption and emission. 
Note the very different flux scales for the various panels.
}
\label{fig cumulative flux}
\end{figure*}

Fig.~\ref{fig cumulative flux} (middle column) shows that
in the model without Razin and without free-free
absorption,
the strongest contribution comes from relatively close
to the apex (up to $\sim$~$300$~$R_{\sun}$, i.e. about $3$ times the binary
separation). This is to be expected as the collision is
strongest close to the apex, so more energy is available for the
acceleration of the particles. On the other hand, inverse Compton
cooling should be strongest there as well, leading to a reduction
of the emission. 

The resulting balance between acceleration and inverse Compton
cooling can be seen in
Fig.~\ref{fig momentum} (lower panel), 
where we plot the maximum momentum ($p_{\rm c}$)
that can be reached due to inverse Compton cooling at the shock where
the electrons are accelerated. We plot this value as a function
of distance along the contact discontinuity ($l_{\rm CD}$). 
For comparison we also plot the 
``characteristic'' momentum for 3.6 and 6 cm emission ($p_6$ and $p_{3.6}$), 
i.e. the momentum at 
which the synchrotron emission is largest for a given wavelength.
Its value can be derived from the fact that maximum emission occurs 
at frequency
$\nu = 0.29 \nu_{\rm s}$ \citep[see, e.g.][Eq.~2.52]{2005PhDT.........1V},
where $\nu_{\rm s}$ is given by Eq.~\ref{eq nu_s}.
Close to the apex, $p_{\rm c}$ is considerably larger than $p_6$ (and
$p_{3.6}$). Because of the (approximate) power-law distribution of the
momentum, there are a large number of electrons available
having their maximum emission at that wavelength. Moving towards the
outer regions, $p_{\rm c}$ becomes closer to, or even lower than,
$p_6$ (and $p_{3.6}$),
thereby explaining the decrease of the emission.

Fig.~\ref{fig results} (middle column) shows that
the maximum flux occurs at apastron (phase 0.5), the minimum at periastron
(phase 0.0). At apastron, the stars are further apart and their winds have
reached the highest speed before they collide. However, this effect cannot
explain why maximum occurs at apastron. The energy available for the 
relativistic electrons is proportional to the shock energy 
$1/2\;\rho \varv^2$, which scales as $\varv/r^2$. 
The higher value for the velocity $\varv$ at apastron
cannot compensate for the lower $1/r^2$ value. The main reason 
for maximum to occur at apastron is the (approximate) scale-invariance
of the colliding-wind region. As the distance between the two stars
becomes larger, the size of the synchrotron emitting region grows
accordingly. Although the synchrotron emissivity is lower at apastron
than at periastron, the integration over the larger size 
of the colliding-wind region results in a higher apastron flux.

\subsubsection{With Razin effect, without free-free absorption}
\label{sect with Razin without free-free}

Next, we switch back on the Razin effect but we still leave out the
free-free mechanism
(Fig.~\ref{fig results} and \ref{fig cumulative flux}, right column).
This results in a substantial
reduction of the flux levels
(e.g.
the 6~cm maximum flux goes down from $\sim$~700~mJy to $\sim$~0.4~mJy).
The large influence of the Razin effect on the flux
can be understood by making a few simple estimates.
The crucial quantity is the factor $f$ defined in Eq.~\ref{eq Razin}:
\begin{equation}
f (\nu,p) = \left[ 1+ \frac{\nu_0^2}{\nu^2} \left( \frac{p}{m_{\rm e} c} \right)^2 \right]^{-1/2} ,
\label{eq Razin f-factor}
\end{equation}
where $\nu$ is the radio frequency, 
$m_{\rm e}$ the mass of the electron,
$c$ the speed of light,
$p$ the momentum
of the electron and
$\nu_0=\sqrt{n_{\rm e} e^2/(\pi m_{\rm e})}$.
In the $\nu_0$ definition, $e$ is the charge of the electron
and $n_{\rm e}$ is the number density of the \emph{thermal} electrons
(i.e. those of the non-relativistic plasma).
The $f$-factor takes on values between 0 and 1 with
$f=1$ indicating no effect, and $f \ll 1$
indicating a strong Razin effect. Taking the frequency at which the electron
has its maximum emission, this
translates into $\nu \ll \nu_{\rm R}$, where
\begin{equation}
\nu_{\rm R} = 20 \frac{n_{\rm e}}{B}
\end{equation}
\citep[see also][Eq.~2.53]{2005PhDT.........1V}. We use 
Eq.~\ref{eq magnetic field simple} for the magnetic field $B$ and
Eq.~\ref{eq ne} to relate $n_{\rm e}$ to the mass density $\rho$. Determining
$\rho$ from the mass-loss rate and approximating
the velocity $\varv$ by the
terminal velocity $\varv_\infty$ results in a $r^{-1}$ scaling of
$\nu_{\rm R}$. Putting in the values for the primary
(Table~\ref{table parameters})
and using
for $\nu_{\rm R}$ the frequency corresponding to 6~cm, 
we find a critical
radius of $\sim$~$1100$ $R_{\sun}$. The Razin effect is important
below that radius, which is indeed where we see the largest effect
in the model calculations (Fig.~\ref{fig cumulative flux}, 
comparing middle and right column).

In this model without free-free absorption, we note that the
Razin effect
does not change the position of maximum flux, which remains at phase 0.5.
It does have an important effect on the
spectral index, which becomes clearly positive ($\ga +1.9$).
The large change in the spectral index can be explained by the
frequency dependence of the Razin effect.
The frequency dependence of $f$ 
(Eq.~\ref{eq Razin f-factor})
is such that $f_{\rm 6 cm} < f_{\rm 3.6 cm}$,
so the 6~cm flux is more affected than the 3.6~cm one. 
This leads to a 
strong change in the spectral index. Note that this change is highly
sensitive to $n_{\rm e}$. 
To quantify this sensitivity, we calculate an artificial model 
without free-free absorption where we 
reduce the $n_{\rm e}$ value in the
$\nu_0$ equation by a factor 30.
This leads to a spectral index that is below 0.2 during
that part of the orbit which is near flux maximum.
We recall that the values used for $n_{\rm e}$ in our model 
(Eq.~\ref{eq ne})
are not based on hydrodynamics. The true values could therefore be very
different, which will
certainly influence the absolute fluxes as well as the spectral index.

The variability seen in Fig.~\ref{fig results} (right column) is
phase-locked, with minimum flux occurring at periastron. Without the
Razin effect (see Sect.~\ref{sect without Razin without free-free}),
this is mainly due to the smaller size of the synchrotron emitting
region. As pointed out above, the Razin effect is more important
in regions of higher density (of the non-relativistic plasma) and
will therefore reduce the flux at periastron more than
at apastron. This explains why the minimum still occurs at
periastron (Fig.~\ref{fig results}, right column) and why the
flux contrast between apastron and periastron is higher than
in the model without Razin effect (Fig.~\ref{fig results}, middle column).

\subsubsection{With Razin effect, with free-free absorption}
\label{sect with Razin with free-free}

In the final step, we again switch on the free-free absorption and emission
to obtain the results in Fig.~\ref{fig results}, left column. The 
absorption 
decreases the flux levels, but it does not introduce a shift
in maximum or minimum phase
(this is discussed further in Sect.~\ref{section improved model}).
The spectral index is now somewhat less positive
compared to the model without free-free absorption. 
This shows that both the Razin
effect and free-free absorption influence the observed spectral index.

\begin{figure}
\resizebox{\hsize}{!}{\includegraphics{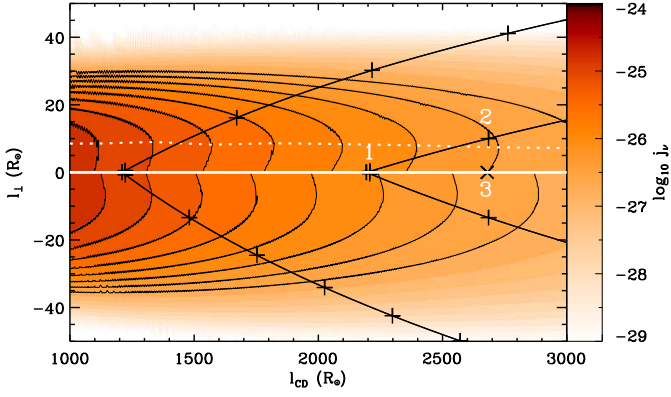}}
\caption{
Part of the 6~cm synchrotron emission region, plotted
as a function of $l_{\rm CD}$ (the length along the contact discontinuity)
and $l_\bot$ (the distance from the contact discontinuity).
The colour scale and contour lines
indicate the value of the emissivity ($j_\nu$). 
The solid white line gives the position of the contact discontinuity 
and the shocks (which we assume to coincide). 
Positive $l_\bot$ values are oriented towards the primary.
Because of the different physical conditions on either side of the
contact discontinuity there is also a discontinuity in $j_\nu$.
The dotted white line shows the offset of maximum emissivity from
the shock (for the side towards the primary).
For a selected number 
of points along the shock, we plot the tracks of the relativistic electrons 
as they are accelerated and advect out and cool down. These tracks
are indicated by the black solid lines. The `+' symbols on each track indicate
one out of every 10 time steps we took. 
The positions indicated by the numbers $1-3$ are discussed in the text
(Sect.~\ref{sect with Razin with free-free}).
        }
\label{fig synchrotron emission}
\end{figure}

A more detailed look at the emission (Fig.~\ref{fig synchrotron emission})
reveals that most of the flux
we detect is formed at some distance from the shock and must therefore
be due to electrons that have cooled down somewhat.
The figure shows 
the synchrotron emissivity $j_\nu$ in a coordinate system linked to the
contact discontinuity.
We limit the figure to distances of
$1000-3000$~$R_{\sun}$, which is where most of the detectable radiation
comes from.

Some care must be taken in the interpretation of this figure. The tracks
show the particles to be moving away from the contact discontinuity/shock
(which we assume to be at the same geometric position). This
differs from the true hydrodynamical situation, where the particles move 
from the shock towards the contact discontinuity. This difference is a
direct consequence of the simplifications in our model. It has a
negligible influence, however, on the resulting fluxes and spectral index
because the thickness of the synchrotron emitting
region is quite small.

For a given point on the contact discontinuity (i.e., for a given
$l_{\rm CD}$),
the highest $j_\nu$ values are found at $l_{\bot} \approx 7 - 8~R_{\sun}$
from the shock (Fig.~\ref{fig synchrotron emission}, dotted white line).
The electrons responsible for this emission
can be traced back to their acceleration
point, which is some 300~$R_{\sun}$ earlier (in $l_{\rm CD}$). 
In addition to the strong $j_\nu$ changes orthogonal to the
contact discontinuity, there
is also a much weaker decreasing gradient as we move from
the inside to the outside. In the flux calculation the outer regions get
higher weight as this emissivity is rotated in 
to a three-dimensional volume.

To understand why the maximum is offset from the shock we 
simplify the situation
by assuming that the electrons emit only at their maximum
frequency and we concentrate on $p_6$, the
characteristic momentum for 6 cm emission.
At the shock at position `1' (Fig.~\ref{fig synchrotron emission}),
a certain number density, $N_1(p_6)$, of relativistic electrons is
responsible for the 6~cm emission. Neglecting a weak function of $p_{\rm c}$,
this number density is proportional to $\rho_1 \Delta u_1^2$
(Eq.~\ref{eq normalization factor}) with the subscript indicating the
position. 
As these electrons advect out to position `2', 
their number density decreases in
proportion to $\rho$. Furthermore, while advecting,
these electrons cool and
therefore another set of electrons becomes
responsible for the 6~cm emission. 
These are electrons that had $p > p_6$ at position `1'; their number
density is less, so we introduce a factor $\epsilon$ ($<1$) to take that
into account. At position `2' their number density
is $N_2(p_6)$, which is 
proportional to $\epsilon\, \rho_2 \Delta u_1^2$. 
Position `3' is 
nearly the same as `2', but it is on the shock. There, new relativistic
electrons are generated with $N_3(p_6) \propto \rho_2 \Delta u_3^2$
(where we used $\rho_2 \approx \rho_3$). 
Whether position `3' or `2' has the highest number density translates in
to a comparison of $\rho_2 \Delta u_3^2$ and 
$\epsilon\, \rho_2 \Delta u_1^2$. The $\Delta u$ gradient is such that
$\Delta u_3 < \Delta u_1$ (Fig.~\ref{fig momentum}, upper panel), so
the result depends on whether the cooling is strong enough to overcome
the effect of the $\Delta u$ gradient. The model calculations
show that this is not the case and therefore $N_2 > N_3$, resulting in
a maximum emission that is offset from the shock.
Only at large distances (beyond those shown on
Fig.~\ref{fig synchrotron emission}),
does the
compensating effect no longer work and there the models show the maximum
to be at the shock. This, however, happens 
beyond the geometric region that contributes significantly
to the observed radio flux.

It may seem surprising that the contribution of the outer regions
of the wind is still detectable, considering that the energy going
into the shock is small (as velocities are nearly tangential
to the contact discontinuity).
We can make a simple estimate to compare the inner and outer wind regions, 
based on the energy available from the shock.
At a position $r$ along the contact discontinuity, we consider a
section of length $\Delta r$.
In a time interval $\Delta t$ a volume of material 
$2\pi r \, \Delta r \, \varv_{\bot} \Delta t$ 
crosses the shock through that section.
The energy input rate is then:
$1/2 \; \rho \Delta u^2 \, 2\pi r \, \Delta r \, \varv_{\bot}, $
of which a fraction $\zeta$ is used to accelerate the particles.
The velocity orthogonal to the shock ($\varv_{\bot}$) is proportional to
$r^{-1}$. 
We derive $\rho$ from
the mass conservation equation in a spherically symmetric wind  and 
approximate $\Delta u$ by $\varv_{\bot}$.
If we assume that some constant fraction of the energy of accelerated
particles goes into the synchrotron emission, we find that
the radio luminosity of section $\Delta r$
is proportional to
$\Delta r/r^4.$
An outer region of size $2000~R_{\sun}$ therefore contributes 
a fraction $10^{-3}$
of an inner region of size $200~R_{\sun}$. If we scale this to the calculated
flux value of $\sim$~700 mJy for the inner region 
(Fig.~\ref{fig cumulative flux},
middle column), we find 0.7~mJy for the outer region. None of the inner
emission is detected due to the Razin effect and free-free absorption.
The resulting flux is therefore predicted to be
$\sim$~0.7~mJy, which is consistent with
the detailed models.

In summary, the standard model shows that phase-locked radio flux variability
is indeed possible in Cyg~OB2 No.~8A. The model fails however in predicting
the phases of maximum/minimum flux and the spectral index.

\subsection{An improved model}
\label{section improved model}

One possible approach to solve the problem 
with the phases of maximum/minimum flux is to ``mirror-reverse" the
system by exchanging
the wind parameters of the primary and secondary star.
It is important to realize that
the star and wind parameters listed in Table~\ref{table parameters}
were not derived directly from information of the binary system, but
rather from typical values for
single stars with the same spectral type as the binary
components. It is not clear how well such a procedure works for 
determining stellar parameters of binary components.
This leads us to a more general approach, where we also explore models 
with star and/or wind parameters that are different from those
listed in Table~\ref{table parameters}.

\begin{figure}
\resizebox{\hsize}{!}{\includegraphics{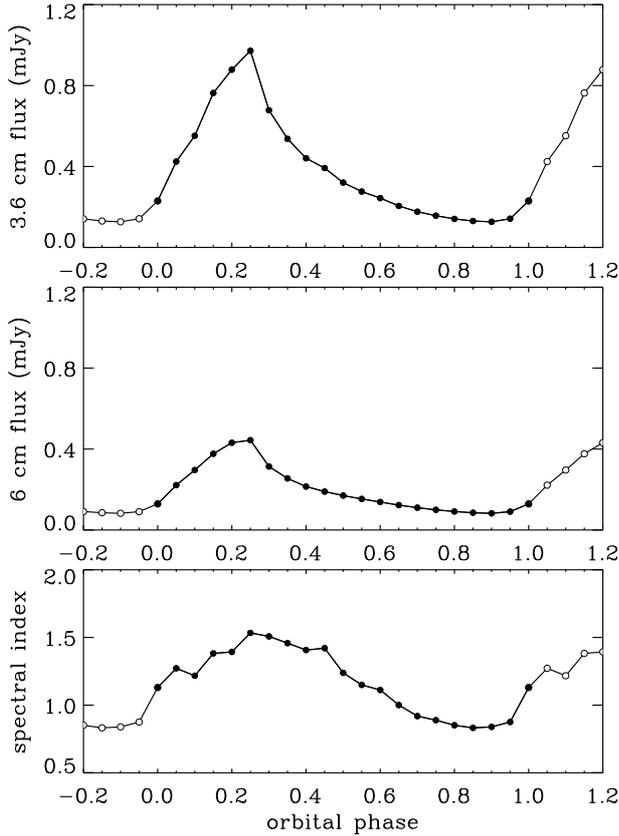}}
\caption{Results for the improved model plotted as a function of orbital 
phase. In this model
the mass loss rate of the primary is $1.0 \times 10^{-6}$
instead of $4.8 \times 10^{-6}\,M_{\sun}\,{\rm yr}^{-1}$ and the
terminal velocity is 2500 instead of 1873 ${\rm km\,s}^{-1}$.
The upper panel shows
the 3.6 cm flux, the middle panel the 6 cm flux and the lower panel
the spectral index between those two wavelengths. 
As for Fig.~\ref{figure fluxes}, open symbols indicate duplications
in the extended phase range.}
\label{fig non standard model}
\end{figure}

In Fig.~\ref{fig non standard model} we present one possible improved
solution where we change the mass loss rate of the primary from
$4.8 \times 10^{-6}$ to $1.0 \times 10^{-6}\,M_{\sun}\,{\rm yr}^{-1}$
and the terminal velocity from 1873 to 2500 ${\rm km\,s}^{-1}$. 
We present only this model from the large number of possibilities we
explored, not because it presents a best-fit solution,
but rather because it requires only minimal changes to the parameters
compared to our standard model. Note that these changes result
in the primary now having a weaker wind (less ram pressure)
than the secondary.

\begin{figure}
\resizebox{\hsize}{!}{\includegraphics{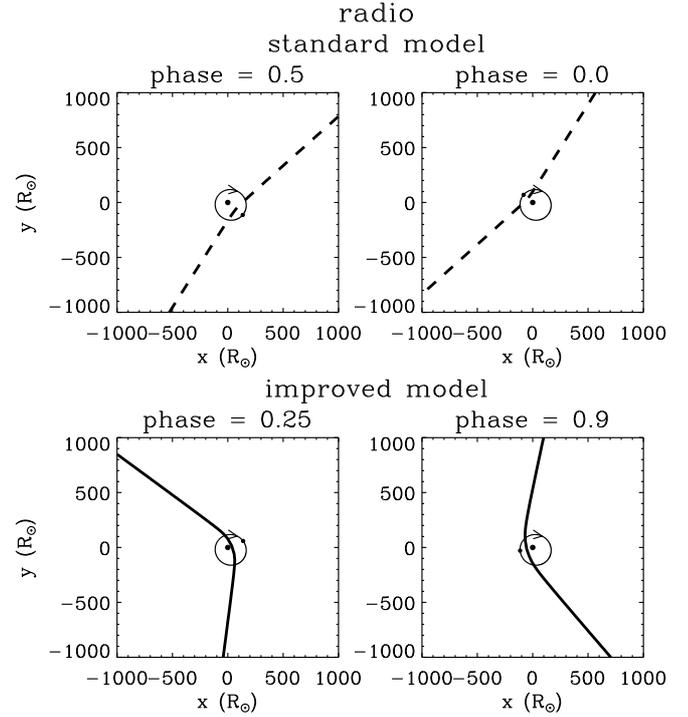}}
\caption{Contact discontinuity of the standard model (top) and 
the improved model (bottom) plotted in the orbital plane. 
For both models, the configuration
is shown for maximum (left) and minimum (right) flux.
Also indicated are the primary and secondary stars, as well as the
binary orbit (assumed centred on the primary). 
The observer looks from the bottom of the page towards the top.
}
\label{fig CDs}
\end{figure}

We see in Fig.~\ref{fig non standard model} that the phase of maximum
is now 0.25, which is closer to the observed value (phase 0.1).
Also the position of minimum flux
at phase 0.9 (observed: 0.6) is improved compared to the
standard model. The main reason for this is the change in shape of the
contact discontinuity as can be seen in Fig.~\ref{fig CDs}.
The upper part of Fig.~\ref{fig CDs} shows the position of the contact
discontinuity for the standard model. The observer looks from the 
bottom of the figure towards the top. We neglect in this discussion
that the observer is not in the orbital plane, but 
58\degr~($=90\degr-i$) above it.
For the standard model we have maximum flux at phase 0.5 
(Fig.~\ref{fig CDs}, top left) because one arm of the contact discontinuity
is then pointing in the direction of the observer and the star with the
weaker wind (the secondary) is in front, so there is only a limited
amount of absorption. At minimum (Fig.~\ref{fig CDs}, top right)
the contact discontinuity is well hidden by the stronger wind from the
primary, which is in front.

The situation is quite different in the improved model. The secondary
now has the stronger wind and therefore
the contact discontinuity wraps around the primary. Maximum flux
(Fig.~\ref{fig CDs}, bottom left)
again occurs when one arm of the contact discontinuity
is pointing in the direction of the observer and the star with the
weaker wind (primary) is in front, but this now happens at phase 0.25.
At phase 0.5, the intrinsic synchrotron emission is even higher
(see discussion Sect.~\ref{sect with Razin without free-free}), but the
arm has rotated further along the orbit so that 
most of it is absorbed by the free-free absorption.
At phase 0.9 (Fig.~\ref{fig CDs}, bottom right) the contact 
discontinuity is hidden by the stronger wind from the secondary 
and therefore the flux is minimum.

Although the phases of minimum and maximum are now closer to their
observed values, the agreement is still not very good. However,
the present model does not take into account the effect of orbital
motion. The importance of this effect is shown by, e.g.
\cite{2009MNRAS.396.1743P}. 
In the specific case of Cyg~OB2 No.~8A, we have 
significant emission coming from $10 - 30$ times the binary separation;
orbital effects at these distances will definitely play a role.
Because of the orbital motion,
the contact discontinuity will get distorted resulting in a
leading arm that is further from the strong-wind secondary than
the trailing arm. 
This will shift the phases at which minimum and maximum occur.
Consider, e.g. maximum flux, where the 
part of the contact discontinuity closest to the observer
is pointing in the direction of the observer (phase 0.25 in Fig.~\ref{fig CDs},
bottom left).
Including orbital motion will push this further
away along the orbit from the strong-wind secondary.
In order to have the contact discontinuity pointing again
in the direction of the observer, we need to move the secondary
back somewhat in its orbit. Maximum flux will thus occur 
at an earlier phase. Including orbital motion will therefore
result in a better agreement with the observed phases
of maximum and minimum flux. 

One remaining problem is the spectral index, which is still 
considerably different from the observed value of $\sim$~0.0.
The spectral index is the result of the combined influence of
free-free absorption and the Razin effect.
As we saw in Sect.~\ref{sect with Razin without free-free},
the spectral
index is highly sensitive to the value of the thermal electron
density $n_{\rm e}$, through the
Razin effect. The physical basis of this effect is that the
local thermal electrons result in a refraction index which changes the
relativistic beaming of the synchrotron radiation
thereby reducing the emitted power
\citep{1965ARA&A...3..297G}. 
In Sect.~\ref{sect with Razin with free-free} we showed that introducing
free-free absorption changes the spectral index as well.

In order to model the observed spectral index, the simplest possibility
is to reduce the mass-loss rates of the stars. 
The reduced density will result in both less free-free absorption and
a reduction of the Razin effect.
As synchrotron emissivity
is proportional to $n_{\rm e}$ this suggests that
the flux would go down. But this is easily compensated for by the
flux increase due to the reduction of the Razin effect, as 
a comparison of Fig.~\ref{fig results} middle and right columns
shows.
In order to be
consistent with observations of single stars, such a reduced-density wind
will need to be clumped \citep[e.g.][]{2008A&ARv..16..209P}.
This would complicate the proposed explanation: the clumping would
again increase the free-free absorption, probably to levels comparable
to that of a smooth high-density wind. This, in turn, could be
counteracted by the destruction of clumps in colliding-wind binaries
\citep{2007ApJ...660L.141P}.
Another possibility is the effect of porosity
\citep{1998ApJ...494L.193S, 2006ApJ...648..565O}, where at least some of
the clumps are optically thick thereby allowing more radiation to escape
through the inter-clump medium.
A further effect is that hot, low-density regions may persist even in the 
radiative part of the colliding-wind region 
\citep[especially on the leading edge of each arm,][]{2009MNRAS.396.1743P}.
The low density will reduce the Razin and free-free effects, 
resulting in a spectral
index closer to the observations.

We also recall that, in our model, we left out the free-free absorption
and emission due to the density and temperature changes in the colliding-wind
region. \citet{2010MNRAS.403.1633P} presents models that study this effect.
His Fig.~14 shows that the free-free contribution from the colliding-wind
region has an intrinsic spectral index which can be close to 0. However,
including the free-free contribution of the winds results in a positive
index. This therefore does not help in explaining the observed
spectral index.

\begin{figure}
\resizebox{\hsize}{!}{\includegraphics{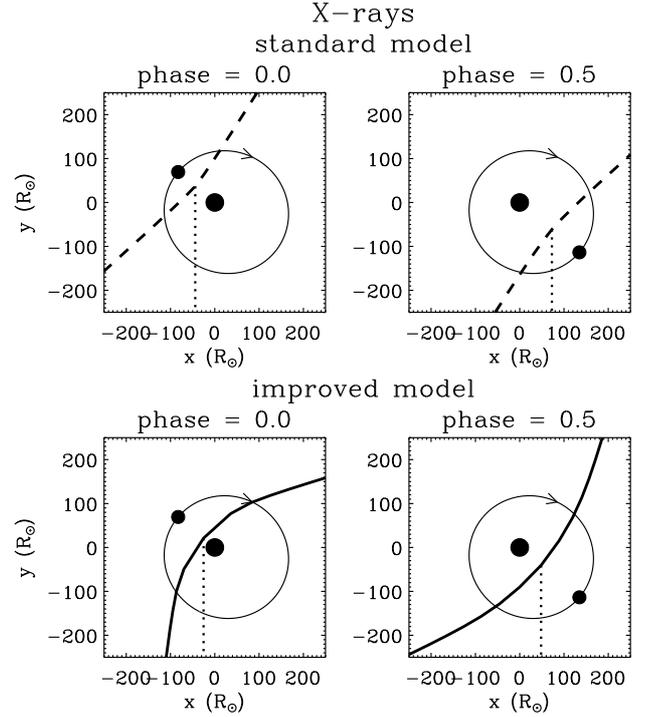}}
\caption{Contact discontinuity of the standard model (top) and 
the improved model (bottom) plotted in the orbital plane. 
For both models, the configuration
is shown for periastron (left) and apastron (right).
Primary and secondary star, and binary orbit are as indicated in
Fig.~\ref{fig CDs}. The dotted line shows the sight-line from the
observer to the apex of the contact discontinuity.
}
\label{fig CDXs}
\end{figure}

\subsection{X-ray emission}
\label{section X-ray emission}

In Sect.~\ref{sect X-ray data}, we noted the anti-correlated behaviour
of the X-ray and radio light curves. The observed X-ray flux is caused by
thermal emission due to heating in the wind-collision zone and
subsequent absorption in the stellar wind material. Both the heating
and the absorption change with stellar phase and thus lead to variability,
just as for the radio fluxes. But the difference with the radio emission
is that the optical depths are very much smaller in X-rays.
For single stars, thermal
X-ray emission that is formed in shocks due to the radiative
driving instability can be seen down to $1-2$ stellar radii
\citep{2003ApJ...592..532K}. Radio emission, on the other hand, can only
be seen from $\sim 100$~ stellar radii. In a massive colliding-wind
binary such as Cyg~OB2 No.~8A this means that the radio emission will
come from outer parts of the contact discontinuity, but the X-ray
emission will be formed close to the apex of the 
contact discontinuity.

Minimum flux occurs at periastron in this eccentric binary, where
the distance between the two components is smallest. The winds
have attained only a fraction of their terminal velocities before
colliding,
resulting in a lower post-shock temperature. In addition, with the reduced
separation, the X-ray emitting plasma is buried in denser wind material.
This leads to a stronger absorption of the X-rays emitted by the plasma
heated by the colliding winds, which is clearly seen in the measured
local absorption column densities (Fig.~\ref{fig nwind}, upper panel).
This results in a strong decrease of the
emerging X-ray flux at this orbital phase, even though an increase of
intrinsic X-ray flux (that scales with the emission measure) is expected.
At apastron, the winds have higher velocities before colliding, resulting
in a higher post-shock temperature, but the absorption by the wind
material surrounding the apex of the wind interaction region (where the
bulk of the X-rays are produced) is significantly reduced. The global
shape of the X-ray light curve results in fact from the combined effect of
varying post-shock temperature as a function of the separation, changing
photoelectric absorption along the line of sight, and decreasing emission
measure when going from periastron to apastron.

When we replace the standard model by the improved one, the contact
discontinuity now wraps around the primary rather than
the secondary. This changes the absolute position of the apex, but
not the relative behaviour of the X-ray emission measure
as a function of orbital phase.
Phases of minimum and maximum emission measure
will be the same for the two models.

Fig.~\ref{fig nwind} (upper panel) 
shows that the observed absorption column changes with
orbital phase. This puts an additional constraint on the models.
To see if our modelling is qualitatively consistent with this, we 
consider the sight-line of the observer toward the apex
(Fig.~\ref{fig CDXs}) and we discuss the absorption column
along that line. At periastron (phase 0.0)
in the standard model, the sight-line passes through the stronger wind 
while at apastron (phase 0.5) it passes through the weaker wind.
Because of the position of the apex, the sight-line
also passes closer to the star at periastron than at apastron
(Fig.~\ref{fig CDXs}, upper panel). One should also take into account
that the sight-line is not in the plane of the orbit, but at an
angle of 58\degr~($=90\degr-i$) above it. For the standard model,
maximum absorption therefore occurs near periastron, minimum absorption
near apastron, which is consistent with the observed behaviour
shown in Fig.~\ref{fig nwind} (upper panel).
For the improved model, on the contrary,
the sight-line goes through the weaker wind at periastron.
This might suggest a reversal of the phases for minimum and maximum
absorption. However, the dominant effect is how close the sight-line
passes to the relevant star. Fig.~\ref{fig CDXs} (lower panel) clearly
shows that this still occurs at periastron. Both the
standard and improved model are therefore qualitatively in agreement
with the observed behaviour of the absorption column.

\cite{2010MNRAS.403.1657P} present models for the thermal X-ray emission
in O+O star binaries. They discuss Cyg~OB2 No.~8A, although no
specific model for this system is presented. One of their models is
an eccentric binary which has a light curve qualitatively similar
to the lower panel in
Fig.~\ref{figure fluxes}: it shows a roughly constant flux for a
large part of the orbit, followed by a rather sharp maximum which in
turn is followed by the minimum. Also their model predicts changes
in the hottest component of the emission, which is indeed observed,
though caution should be used in the interpretation of 
this result as the error bars are quite large
(see Sect.~\ref{sect X-ray data}).

\subsection{Future work}

The improved model presented here is qualitatively successful in explaining a number
of important features of the radio emission of
the colliding-wind binary Cyg~OB2 No.~8A.
Nevertheless, it should be realized that some important physical
ingredients are still missing. 
The major ones are including orbital motion and solving
the equations of hydrodynamics. Orbital motion will have an important effect
on the phases of minimum and maximum flux.
Solving the hydrodynamics equations
will allow us to correctly position the
shocks and take into account the instabilities and inhomogeneities,
as they influence both the thermal free-free emission and absorption,
and the synchrotron emission.
Information from hydrodynamics will also
allow us to replace the assumed shock strength
$\chi=4$ by its correct value.
This will influence
the relativistic electron momentum distribution. In addition, clumps that
are optically thick at radio wavelengths will result in porosity,
letting a larger fraction of the synchrotron emission escape.
Finally, future models should handle both the X-ray and radio emission
simultaneously,
giving us a consistent picture of these colliding-wind systems.

\section{Conclusions}
\label{sect conclusions}

We determined the radio light curve of Cyg~OB2 No.~8A, based on new and
archive VLA observations. We find phase-locked variability in the radio
fluxes. This is contrary to expectations, as the synchrotron emission
due to the colliding winds should be hidden by free-free absorption
in this close binary. We also present an updated version of the X-ray
light curve. The radio and X-ray light curves are nearly anti-correlated,
which is another unexpected result.

Our model calculation for the radio flux shows that the synchrotron 
emission region
extends considerably beyond the apex of the colliding-wind region,
explaining why we do see phase-locked variability.
We achieve this result without any need for clumping or porosity
in the model.
Our standard model -- which uses the star and wind parameters
proposed by \cite{2006MNRAS.371.1280D} -- does not correctly predict
the phases of minimum and maximum. When we change the wind parameters
(which are not well-determined) to make the secondary have the stronger
wind, we find a considerable improvement in phases of minimum/maximum.
A more detailed study of Cyg~OB2 No.~8A is therefore needed
to better determine the star and wind parameters of its binary components.
To improve the phase agreement even further we will need to include orbital
motion in the model.

The nearly anti-correlated behaviour of the X-ray and radio light
curves is due to their very different formation regions. X-rays are
formed quite close to the apex of the contact discontinuity, where
the heating is strongest and they suffer only a limited amount of
absorption. Radio emission, on the other hand, is strongly absorbed, so
the radio emission we can detect is
formed much further out along the contact discontinuity. Because the
contact discontinuity is curved, this results in X-rays and radio having
their minimum/maximum flux at very different phases.

The model values for the radio spectral index 
(+0.75 to +1.5)
differ considerably from the observed value of $\sim$~0.0. 
However, the model values are quite
sensitive to the density of the thermal electrons, which are responsible
for the Razin effect and the free-free absorption.
A decrease of that density would allow a better
agreement with the observations. Such a decrease could be due to 
lower density regions caused by hydrodynamical effects
in the colliding-wind region or by clumping in a wind
with a reduced mass-loss rate.
Although we did not require clumping or porosity to explain the
phase-locked variability, it is likely that we will need it to explain 
the observed spectral index.
Because of the high sensitivity of the spectral index to clumping conditions,
it could provide a valuable pathway to investigate the well-known problem
of clumping in single-star winds.
Future, improved models, 
applied to Cyg~OB2 No.~8A and other non-thermal radio emitters
should therefore
be capable of determining limits on this clumping.

\begin{acknowledgements}
We thank Joan Vandekerckhove for his help with the reduction of the
VLA data. We are grateful to
the original observers of the VLA archive data used in this paper.
We also thank Patrick M\"uller for discussions about the numerical code.
We thank the referee for comments that helped to improve this paper.
D. Volpi acknowledges funding by the Belgian Federal Science Policy Office 
(Belspo), under contract MO/33/024.
This research has made use of the SIMBAD database, operated at CDS,
Strasbourg, France and NASA's Astrophysics Data System Abstract Service.
\end{acknowledgements}

\bibliographystyle{aa}
\bibliography{cyg8A}

\Online
\appendix

\section{Modelling}
\label{appendix modelling}

Most of the calculations discussed here are limited to the 
two-dimensional (2D) plane of the binary orbit. Only in 
Sect.~\ref{sect map into 3D} do we rotate the results of the
2D calculations into the three-dimensional (3D) simulation volume that is used
to calculate the resulting flux.

\subsection{Binary orbit}

\begin{figure}
\resizebox{\hsize}{!}{\includegraphics{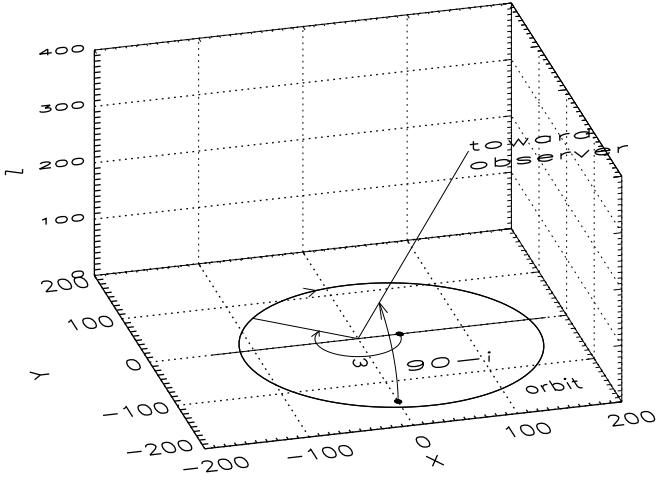}}
\caption{Binary orbit in the XY plane of
the 3D simulation volume. The primary is positioned
at the centre of the volume. The argument of periastron is denoted $\omega$
and the inclination angle $i$. The units on the axes are
$R_{\sun}$.
        }
\label{fig orbit}
\end{figure}

Figure~\ref{fig orbit} shows how the
two stars are positioned in the XY plane of the 3D
simulation volume.
The primary is at the centre of the volume
and the position of the secondary is
determined by the orbital phase. 
The angle of periastron ($\omega$) is measured from
the intersection of the plane of the sky with the
plane of the orbit. The sight-line towards the observer is in the
YZ plane and the inclination angle $i$ is the angle between the
plane of the sky and the orbital plane. The secondary rotates
clockwise in this figure (as seen from above). Periastron
corresponds to phase 0.0.

\subsection{Contact discontinuity}

The position of the contact discontinuity at any given orbital phase
is determined by the balance of ram pressure between the velocity components
orthogonal to
the contact discontinuity \citep[see equations in][]{2004ApJ...611..434A}.
These equations take into account that the terminal velocity of the winds
may not have been reached yet before they collide.
Specifically, in the Cyg~OB2 No.~8A standard model 
presented in Sect.~\ref{section standard model},
the winds collide with $0.65 \, \varv_\infty$ (primary) and 
$0.71 \, \varv_\infty$ (secondary) at the apex at periastron. 
After the stellar wind material has passed through the shock, it
then moves outward parallel to the contact discontinuity
(there is also a small orthogonal velocity component, see 
      Sect.~\ref{sect advection and cooling}).
We assume that the shocks and the contact
discontinuity are sufficiently close to one another that we do not
need to make a distinction in their geometric positions.
On each side of the contact discontinuity, we assume the stellar wind
to have a $\beta=1$ velocity law and a density derived from 
the mass conservation law, using the assumed velocity law
and mass-loss rate.

\subsection{Momentum distribution relativistic electrons}

We determine the momentum distribution of the relativistic electrons 
for a grid of points
along the two shocks. We start with a population of electrons
generated at a given point on the shock itself. We then follow the
time evolution of this population as it advects away and cools down.
The momentum distribution uses a grid that is uniform in
$\log p$, between $p_0 = 1.0$ MeV/$c$ 
(for the choice of this specific value, see
     \citet{2004A&A...418..717V})
and the upper limit $p_{\rm c}$ (defined below, Eq.~\ref{eq pc})
These limits are sufficiently wide to
cover all of the radio emission we are interested in.
We stop following the electrons when their 
momentum falls below $p_0$, or when they leave the simulation
volume. (In our calculations we checked that the
emission beyond the simulation volume is negligible.)

\defcitealias{2005PhDT.........1V}{VL}
In the following, we refer mainly to 
\citet[][ hereafter \citetalias{2005PhDT.........1V}]{2005PhDT.........1V}
as the source for the equations. Citations to the original papers
can be found in that reference.

\subsection{New relativistic electrons}

At the shock, new relativistic electrons are put into the system.
Taking into account the particle acceleration mechanism
and the inverse Compton
cooling, it can be shown that
their momenta follow a modified power-law distribution 
\citepalias[][Eqs.~5.5, 5.6]{2005PhDT.........1V}:
\begin{eqnarray}
N(p) & = & N_{\rm E} p^{-n} \left(1 - \frac{p^2}{p_{\rm c}^2} \right)^{(n-3)/2} , \\
n & = & \frac{\chi +2 }{\chi -1} .
\end{eqnarray}
With \citepalias[][Eqs. 5.1-3]{2005PhDT.........1V}:
\begin{equation}
\left( \frac{p_{\rm c}}{m_{\rm e} c} \right)^2 = 
\frac{\chi \Delta u^2}{\chi^2 - 1} \frac{4\pi r^2}{\sigma_{\rm T} L_*}
\frac{eB(r)}{c} ,
\label{eq pc}
\end{equation}
with
$m_{\rm e}$ and $e$ the mass and charge of the electron,
$\sigma_{\rm T}$ the Thomson electron scattering cross section,
$c$ the speed of light,
$L_*$ the luminosity of the star
and $r$ the distance of the point on the shock to the star. 
Because there are
two stars, we should calculate the contribution of both:
\begin{equation}
\frac{L_*}{r^2} = \frac{L_{*,1}}{r_1^2} + \frac{L_{*,2}}{r_2^2} ,
\end{equation}
with $r_1$ the distance to the primary and $r_2$ the distance to
the secondary.
Also \citepalias[][Eq.~3.7]{2005PhDT.........1V}:
\begin{eqnarray}
\Delta u = u_1 - u_2 = u_1 \frac{\chi-1}{\chi} ,
\end{eqnarray}
where $u_1$ is the upstream velocity and $u_2$ the downstream one
and the shock strength $\chi=u_1/u_2$.
Throughout this work we assume
strong shocks with $\chi=4$, 
but we note that higher values can be expected in radiative shocks.

The normalization factor $N_{\rm E}$ is related to $\zeta$, the fraction of
energy that gets transferred from the shock to the relativistic particles, by
\citepalias[][Eq.~5.10]{2005PhDT.........1V}:
\begin{eqnarray}
N_{\rm E} = 4 \frac{\zeta}{\displaystyle \log_{\rm e} \left(\frac{p_{\rm c}}{p_0} \right)} \rho_1 \frac{\Delta u^2}{c} \quad ({\rm for}~\chi=4),
\label{eq normalization factor}
\end{eqnarray}
where $\rho_1$ is the upstream density and we 
assume $\zeta=0.05$ \citep{1987PhR...154....1B, 1993ApJ...402..271E}. 

The magnetic field at distance $r$ 
to the relevant star
is assumed to be given by 
(\citealt{1967ApJ...148..217W};
\citetalias[][Eq. 2.48]{2005PhDT.........1V}):
\begin{equation}
B(r) = B_* \frac{\varv_{\rm rot}}{\varv_\infty} \frac{R_*}{r} ,
\label{eq magnetic field simple}
\end{equation}
where $B_*$ is the surface magnetic field (assumed to be 100 G)
and $\varv_{\rm rot}$ is the equatorial rotational velocity.
This equation is only valid at larger distances from
the star, but we use it everywhere
as the region close to the star will not be visible anyway (due to
free-free absorption).
This equation neglects any amplification
of the field (such as described by, e.g.
\citet{2004MNRAS.353..550B})
or any reduction by magnetic reconnection.
For each of the two shocks, the corresponding
magnetic field is taken into account.

\subsection{Advection and cooling}
\label{sect advection and cooling}

During a time step $\Delta t$, the relativistic electrons advect away and
cool mainly
due to inverse Compton and adiabatic cooling 
(see  
\cite{1992PhDT.........1C} and
\citetalias{2005PhDT.........1V} (p.~96)
for a discussion on other cooling mechanisms and why they are not
important in the present situation). 

The equation for cooling of a single relativistic electron
can be written as:
\begin{eqnarray}
\frac{{\rm d}p}{{\rm d}t} & = & -a p^2 +b p \quad\mbox{with} \label{eq cooling}\\
a & = & \frac{\sigma_{\rm T}}{3 \pi m_{\rm e}^2 c^3} \frac{L_*}{r^2} , \\
b & = & \frac{1}{3 \rho} \frac{\Delta \rho}{\Delta t} \label{eq adiabatic cooling},
\end{eqnarray}
where $a p^2$ gives the inverse 
Compton\footnote{The inverse Compton cooling should, in principle, 
   use the Klein-Nishina scattering equation instead of the Thomson one. 
   To test the importance of this, we 
   adapted our code to use the Klein-Nishina equation, following the
   description in \cite{2010NJPh...12c3044S}. The input photons are 
   assumed to come from a graybody with a temperature of 40\,000 K.
   The Klein-Nishina equation leads to a somewhat 
   more complicated expression than Eq.~\ref{eq cooling one electron}
   which can no longer be inverted analytically. We therefore do the
   time integration numerically, as well as the determination of 
   ${\rm d}p_{\rm init}/{\rm d}p$. The resulting fluxes show 
   fractional differences less than $10^{-4}$. We therefore decided to continue
   using the simpler Thomson scattering in the model.
}
cooling 
\citepalias[][Eq.~5.2]{2005PhDT.........1V}
and $b p$ the adiabatic cooling (which can be derived from the fact that
for adiabatic cooling
$p \propto T^{1/2} \propto \rho^{(\gamma-1)/2}$ and using $\gamma=5/3$).

Integration of Eq.~\ref{eq cooling} gives:
\begin{equation}
p  =  \frac{p_{\rm init} \exp (b \Delta t)}{\displaystyle 1 - \frac{a}{b} p_{\rm init} \left( 1 - \exp (b \Delta t) \right)} ,
\label{eq cooling one electron}
\end{equation}
where $p_{\rm init}$ is the momentum at the previous time step.
We can invert this to:
\begin{equation}
p_{\rm init} = \frac{p \exp (-b \Delta t)}{\displaystyle 1 - \frac{a}{b} p \left( 1 -\exp (-b \Delta t) \right)} .
\label{eq p-pinit simple}
\end{equation}
We will also need the following derivative:
\begin{equation}
\frac{{\rm d}p_{\rm init}}{{\rm d}p} = \frac{\exp (-b \Delta t)}{\displaystyle \left[1 - \frac{a}{b} p \left( 1 - \exp (-b \Delta t) \right)\right]^2} .
\end{equation}

We assume that only the momentum 
$p$ changes significantly with time; for the other
variables, we take their value mid-way between the start and the end
of the time step $\Delta t$.
The density $\rho$ we need in Eq.~\ref{eq adiabatic cooling}
is the one between the shock and the contact
discontinuity. We do not have detailed hydrodynamical
models for this, so we simply
approximate it by $\rho = \chi \rho_{\rm s}$ where $\rho_{\rm s}$ is the density
in the smooth wind at the position of the shock.
We stress once again that in these radiative shocks, the post-shock 
density increases rapidly downstream, maybe by several orders of magnitude,
rather than the $\chi=4$ value we use. 

The above equations describe what happens to a single relativistic
electron. The time evolution of the momentum distribution function over
the time step $\Delta t$ is:
\begin{equation}
N(p) = N(p_{\rm init}) \frac{\rho (t+\Delta t)}{\rho (t)} \frac{{\rm d}p_{\rm init}}{{\rm d}p} .
\label{eq p-distribution simple}
\end{equation}
We stop following the electrons when their 
momentum falls below $p_0$, or when they leave the simulation
volume.

In our simplifying assumptions the contact discontinuity
and the shock are at the same geometrical position. Taken literally, this
would result in a synchrotron emitting region that has zero thickness and
zero synchrotron emission. In the true hydrodynamical situation, particles
move through the shock (which is at some distance from the contact
discontinuity) and then move slowly towards the contact discontinuity.
At the same time, they are being advected outward with a velocity
parallel to the contact discontinuity that is much larger than
the corresponding orthogonal velocity. The thickness of the
synchrotron emitting region is set by the combination of these
two velocities.
In the absence of hydrodynamical information, we use the following
procedure to determine these velocities.
The parallel component ($\varv_{\|}$) at the contact discontinuity
is determined by 
projecting the smooth wind velocity vector on to the contact discontinuity. 
At other positions in the wind, we interpolate in the grid of
$\varv_{\|}$ as a function of the y-coordinate.
The orthogonal component ($\varv_{\bot}$) 
at the contact discontinuity is expected to be of the order of 
the orthogonal component of the material coming into the shock, with a minimum
value corresponding to the thermal expansion of the gas. As a 
simplification, we set it exactly equal to the incoming orthogonal component,
and we use the same type of interpolation
as for the parallel velocity to determine it at other positions.
We reverse the orthogonal velocity direction
to have the gas moving away from the contact discontinuity.
Note that this situation is ``inside out" compared to the
true hydrodynamical situation where the particles move towards
the contact discontinuity. The thickness of the synchrotron emitting
region is so small however that this inversion has negligible influence on
our results.

\subsection{Emissivity}
\label{section emissivity}

The equation for the emissivity is given by 
\citepalias[][Eq.~B.1]{2005PhDT.........1V}:
\begin{eqnarray}
j_\nu(r) & = & \frac{1}{4 \pi} \int_{p_0}^{p_{\rm c}} {\rm d}p N(p,r)
\int_0^\pi \frac{{\rm d}\theta}{4\pi} 2\pi \sin \theta
\frac{\sqrt{3} e^3}{m_{\rm e} c^2} \nonumber \\
& & \quad\quad\quad B f(\nu,p) \sin \theta 
F \left(\frac{\nu}{f^3(\nu,p)\nu_{\rm s}(p,r) \sin\theta} \right) ,
\label{equation emissivity}
\end{eqnarray}
With \citepalias[][Eq.~2.51, 2.50]{2005PhDT.........1V}:
\begin{eqnarray}
f (\nu,p) & = & \left[ 1+ \frac{\nu_0^2}{\nu^2} \left( \frac{p}{m_{\rm e} c} \right)^2 \right]^{-1/2} \label{eq Razin} \\
\nu_0 & = & \sqrt{\frac{n_{\rm e} e^2}{\pi m_{\rm e}}} \\
\nu_{\rm s} (p,r) & = & \frac{3}{4 \pi} \frac{eB}{m_{\rm e} c} 
   \left( \frac{p}{m_{\rm e} c} \right)^2 \label{eq nu_s} \\
F(x) & = & \int_x^\infty {\rm d}\eta K_{5/3} (\eta) .
\end{eqnarray}
We use Eq.~\ref{eq magnetic field simple} for the magnetic field.
Note that this is an approximation, as the shock may compress the
magnetic field as well, leading to higher values than given by
Eq.~\ref{eq magnetic field simple}.
The expression for 
$j_\nu(r)$ is evaluated at position
$r$ and frequency $\nu$. The $f$ function corrects for the Razin effect and
$K_{5/3}$ is the modified Bessel function.
To calculate $n_{\rm e}$,
the number density of the \emph{thermal} electrons (i.e. those in
the non-relativistic plasma),
we use:
\begin{equation}
n_{\rm e}=\frac{\chi \rho_{\rm s}}{m_{\rm H}} \frac{1.2}{1.4} ,
\label{eq ne}
\end{equation}
where $\rho_{\rm s}$ is the smooth wind density and $m_{\rm H}$ the mass
of a hydrogen atom. The factor 1.2/1.4 takes into account a 
fully ionized hydrogen+helium solar composition.
In these radiative shocks the post-shock density may increase
by several orders of magnitude rather than by $\chi=4$,
which will influence the value of $n_{\rm e}$. 

To reduce
the computing time we set $\sin \theta =1$ in the $f \sin \theta$ 
factors of Eq.~\ref{equation emissivity} and we also use:
\begin{equation}
\int_0^\pi \frac{{\rm d}\theta}{4\pi} {2\pi \sin \theta} = 1 .
\end{equation}
The $F$ function is precalculated on a grid of $x$-values; the evaluation
of the function then reduces to an interpolation.

\subsection{Map into 3D grid}
\label{sect map into 3D}

Using the equations from the previous sections, we can make a 
2D map of the synchrotron emission
(part of which is shown in Fig.~\ref{fig synchrotron emission}).
Note that the synchrotron emission extends on both sides of the
contact discontinuity, because there is a shock on either side.
Based on the cylindrical symmetry around the line
connecting the two stars (at the given orbital phase),
we then
``rotate" this 2D map into the 3D
simulation volume.

It is important that during this procedure we do not lose part
of the emission due to interpolation or reduced resolution.
To achieve this, we use volume-integrated emissivities.
While doing the 2D calculation we store the volume integrated
emissivity $\overline{j_\nu(r)}$ defined by:
\begin{equation}
\overline{j_\nu(r)} \Delta V = \int_{\Delta V} j_\nu(r) {\rm d} V .
\end{equation}
The volume $\Delta V$ used in this definition is determined as
follows. We consider the 2D surface covered by (1) the step in the
coordinate along the contact discontinuity
and (2) 
the distance travelled by the relativistic electrons 
between two consecutive time steps.
We then rotate this surface around the
line connecting the two stars
over a small angle $\alpha$, thereby defining the emissivity
volume $\Delta V$.

When we have finished the full 2D calculation,
we rotate the 2D plane into the 3D simulation volume,
using steps of angle $\alpha$
(as defined above). 
     For each of the emissivity volumes $\Delta V$ defined above, and for
a given rotation angle, we determine
what fraction of the rotated volume overlaps with the 
cells in the 3D simulation volume. This is done by 
subdividing each dimension of the $\Delta V$ volume into three.
For each of the resulting $3^3$ points 
we determine in which 
3D simulation cell they end up after rotation.
A running sum for that cell is then updated with
the fraction $3^{-3}$ 
of the volume integrated emissivity $\overline{j_\nu}$
of the rotated volume.
At the end of this procedure, the accumulated emission in each
3D simulation cell is divided by its volume, thus obtaining
$j_\nu$ for that cell.

In this procedure we lose some resolution, specifically
around the apex, where the 2D resolution is quite high but where
the 3D resolution is much lower. However, because of the use of 
volume-integrated quantities, we do not lose any of the synchrotron
emission. Furthermore, the apex of the wind is well 
hidden by the free-free absorption
and will therefore not contribute to the resulting radio flux at
the wavelengths we consider. If our model were to be applied to X-rays, 
optical emission or radio
emission at much shorter wavelengths, a higher-resolution 3D grid would 
be required. 

In the 3D model. we also include the thermal free-free opacity and 
emissivity, which is due to the ionized wind material. 
To calculate
these we use the \citet{1975MNRAS.170...41W} equations.
The winds of both stars are assumed to be smooth.
For the Gaunt factor at frequency $\nu$ we use the equation from
\citet{1991ApJ...377..629L}:
\begin{equation}
g_\nu = 9.77 \left( 1 + 0.13 \log_{10} \frac{T_{\rm e}^{3/2}}{Z \nu} \right),
\end{equation}
where $T_{\rm e}$ is the thermal electron temperature in the wind
and $Z$ the charge of the ions.
The stars themselves are introduced into the simulation
volume as high opacity spheres.
We then solve the radiative transfer equation in Cartesian coordinates
along the line of sight. We use
Adam's \citep{1990A&A...240..541A} finite volume method because of its
simplicity. By repeating the whole procedure for a number of orbital
phases and wavelengths, we calculate the radio light curve at these
wavelengths and determine how the fluxes and spectral index change
with phase.

\section{Data reduction}
\label{appendix data reduction}

The data reduction is accomplished using the Astronomical Image
Processing System (AIPS), developed by the NRAO. We follow
the standard procedures of antenna gain calibration, absolute flux
calibration, imaging and deconvolution. Where possible, we apply
self-calibration to the observations (see notes to 
Table~\ref{table radio data}). We measure the fluxes and error bars by fitting
elliptical Gaussians to the sources on the images. The error bars listed
in Table~\ref{table radio data} include not only the rms noise
in the map, but also an estimate of the systematic errors that 
were evaluated using a jackknife technique. This technique drops
part of the observed data and redetermines the fluxes, giving some
indication of systematic errors that can be present.
Upper limits are three
times the uncertainty as derived above. 
For details of the reduction, we refer
to the previous papers in this series 
\citep{2005A&A...436.1033B, 2007A&A...464..701B, 2008A&A...483..585V}.
We exclude from Table~\ref{table radio data}
those observations with upper limits higher than 6 mJy (which
corresponds to about 3 times the highest detection at all wavelengths).

A comparison with values previously published in the literature
shows that in most cases the error bars overlap with ours.
There are just a few exceptions, which we discuss further. 
\citet{1989ApJ...340..518B}
list only an upper limit for the AC116 2~cm observation, while we find
a detection (though marginal at just above 3 sigma). Our detection
is quite consistent in all variants we consider in the jackknife test
and is at the exact
position of Cyg~OB2 No.~8A, so we consider it a detection.
For the AS483 6~cm, \citet{1998ApJS..118..217W}
find a value of 1.04 $\pm$ 0.08 mJy, which is
$\sim$~20\% lower than our value.
A possible explanation is that they did not correct for the
decreasing sensitivity of the primary beam. This effect is 
important when the target is not in the centre of the image, and
is approximately
20\% for the position of No.~8A on the image.
\citet{2006A&A...454..625P}
list an upper limit of 0.54 mJy for the AS786 6~cm observation,
but we clearly detect Cyg~OB2 No.~8A on the image we made.
\citet{1989ApJ...340..518B}
find a slightly lower value of 1.0 $\pm$ 0.1 mJy for the
AC116 1984-12-21, 20 cm observation.
We have no explanation for these last two discrepancies.

We also take the opportunity to correct some clerical errors
in the literature:
\citet{1989ApJ...340..518B} date the CHUR observations 
as 1980-Mar-22, but it should be 1980-May-23.
The \citet{1998ApJS..118..217W}
1991 observation was made at 3.6~cm, not at 6~cm
and the 1992 observation was made in January, not June.

\begin{longtable}{lllrrcrrrlcl}
\caption{\label{table radio data}Reduction of the Cyg~OB2 No.~8A VLA data. 
}\\
\hline\hline
& \multicolumn{1}{c}{(1)} & \multicolumn{1}{c}{(2)} & \multicolumn{1}{c}{(3)} &
\multicolumn{1}{c}{(4)} & \multicolumn{1}{c}{(5)} & \multicolumn{1}{c}{(6)} &
\multicolumn{1}{c}{(7)} & \multicolumn{1}{c}{(8)} & \multicolumn{1}{c}{(9)} &
\multicolumn{1}{c}{(10)} & \multicolumn{1}{l}{(11)} \\
& \multicolumn{1}{c}{Progr.} & \multicolumn{1}{c}{Date} &
\multicolumn{1}{c}{Ctr.} &
\multicolumn{3}{c}{Phase Calibrator} &
\multicolumn{1}{c}{Int.} &
\multicolumn{1}{c}{No.} &
\multicolumn{1}{c}{Config.} &
\multicolumn{1}{c}{Flux} &
\multicolumn{1}{l}{Notes}\\
\cline{5-7}
&  &  & &
\multicolumn{1}{c}{Name} &
\multicolumn{1}{c}{Flux} &
\multicolumn{1}{c}{Dist.} &
\multicolumn{1}{c}{Time} &
\multicolumn{1}{c}{Ants.} & &
\multicolumn{1}{c}{} & \\
&  &  & &
\multicolumn{1}{c}{} &
\multicolumn{1}{c}{(Jy)} &
\multicolumn{1}{c}{(degr.)} &
\multicolumn{1}{c}{(min)} &
\multicolumn{1}{c}{} & &
\multicolumn{1}{c}{(mJy)} & \\
\hline
\endfirsthead
\caption{continued.}\\
\hline\hline
& \multicolumn{1}{c}{(1)} & \multicolumn{1}{c}{(2)} & \multicolumn{1}{c}{(3)} &
\multicolumn{1}{c}{(4)} & \multicolumn{1}{c}{(5)} & \multicolumn{1}{c}{(6)} &
\multicolumn{1}{c}{(7)} & \multicolumn{1}{c}{(8)} & \multicolumn{1}{c}{(9)} &
\multicolumn{1}{c}{(10)} & \multicolumn{1}{l}{(11)} \\
& \multicolumn{1}{c}{Progr.} & \multicolumn{1}{c}{Date} &
\multicolumn{1}{c}{Ctr.} &
\multicolumn{3}{c}{Phase Calibrator} &
\multicolumn{1}{c}{Int.} &
\multicolumn{1}{c}{No.} &
\multicolumn{1}{c}{Config.} &
\multicolumn{1}{c}{Flux} &
\multicolumn{1}{l}{Notes}\\
\cline{5-7}
&  &  & &
\multicolumn{1}{c}{Name} &
\multicolumn{1}{c}{Flux} &
\multicolumn{1}{c}{Dist.} &
\multicolumn{1}{c}{Time} &
\multicolumn{1}{c}{Ants.} & &
\multicolumn{1}{c}{} & \\
&  &  & &
\multicolumn{1}{c}{} &
\multicolumn{1}{c}{(Jy)} &
\multicolumn{1}{c}{(degr.)} &
\multicolumn{1}{c}{(min)} &
\multicolumn{1}{c}{} & &
\multicolumn{1}{c}{(mJy)} & \\
\hline
\endhead
\hline
\endfoot
\endlastfoot
\multicolumn{3}{l}{\bf 2   cm} \\*
& AA28   & 1984-03-04 & 8 & 2007+404 & 3.33  $\pm$ 0.28  & 4.9  &  20 & 27 & CnB & 0.7   $\pm$ 0.4   & \tablefootmark{a} \\
& AC116  & 1984-12-21 & 8 & 2007+404 & 3.96  $\pm$ 0.07  & 4.9  &  10 & 27 &  A  & 0.62  $\pm$ 0.19  & \tablefootmark{a} \\
& AS786  & 2004-02-15 & 7 & 2015+371 & 2.88  $\pm$ 0.03  & 5.4  &  60 & 26 & CnB &         $<$ 4.    & \\
\multicolumn{3}{l}{\bf 3.6 cm} \\*
& AS397  & 1990-02-16 & 8 & 2007+404 & 3.10  $\pm$ 0.04  & 4.9  &  59 & 22 &  A  & 0.48  $\pm$ 0.05  &  \\
& AH395  & 1990-02-17 & 8 & 2007+404 & 3.05  $\pm$ 0.01  & 4.9  &  44 & 22 &  A  & 0.66  $\pm$ 0.04  &  \\
& AW288  & 1991-07-04 & 8 & 2025+337 & 2.08  $\pm$ 0.06  & 7.7  &   5 & 26 &  A  & 0.70  $\pm$ 0.09  &  \tablefootmark{b} \\
& AW304  & 1992-01-24 & 8 & 2007+404 & 2.87  $\pm$ 0.03  & 4.9  &  12 & 24 & CnB & 1.06  $\pm$ 0.07  &  S \tablefootmark{b}  \\
& BP1    & 1992-05-30 & 9 & 2052+365 & 1.87  $\pm$ 0.01  & 6.0  &  99 & 23 & DnC &         $<$ 1.1   &  \\
& AB671  & 1993-01-24 & 9 & 2007+404 & 3.33  $\pm$ 0.07  & 4.9  &  23 & 27 &  A  &         $<$ 5.    &  S  \\
& AB671  & 1993-01-29 & 9 & 2007+404 & 3.33  $\pm$ 0.02  & 4.9  &  24 & 27 & BnA &         $<$ 2.0   &  S  \\
& AB671  & 1993-02-01 & 9 & 2007+404 & 3.29  $\pm$ 0.03  & 4.9  &   6 & 27 & BnA &         $<$ 4.    &  \\
& AB671  & 1993-02-14 & 9 & 2007+404 & 3.41  $\pm$ 0.03  & 4.9  &   7 & 27 & BnA &         $<$ 4.    &  S  \\
& AB671  & 1993-02-20 & 9 & 2007+404 & 3.8   $\pm$ 0.1   & 4.9  &  14 & 24 &  B  &         $<$ 5.    &  \\
& AS483  & 1993-05-01 & 98& 2007+404 & 3.28  $\pm$ 0.01  & 4.9  &  25 & 27 &  B  & 1.19  $\pm$ 0.13  &  S \tablefootmark{b} \\
& AR277  & 1994-04-17 & 9 & 2007+404 & 2.97  $\pm$ 0.01  & 4.9  &  23 & 14 &  A  &         $<$ 5.    &  S  \\
& AR328  & 1995-04-27 & 9 & 2007+404 & 2.77  $\pm$ 0.01  & 4.9  &  11 & 16 &  D  &         $<$ 2.7   &  \\
& AS644  & 1999-02-05 & 7 & 2007+404 & 2.07  $\pm$ 0.12  & 4.9  &  35 & 23 & DnC & 1.0   $\pm$ 0.3   &  \\
& AS644  & 1999-04-18 & 8 & 2015+371 & 2.17  $\pm$ 0.01  & 4.9  &   5 & 27 &  D  & 0.92  $\pm$ 0.10  &  \\
& AW515  & 1999-06-08 & 9 & 2007+404 & 2.02  $\pm$ 0.01  & 4.9  &  10 & 19 & AD  &         $<$ 2.8   &  I,S  \\
& BB116  & 1999-12-04 & 9 & 2007+404 & 2.20  $\pm$ 0.01  & 4.9  & 101 & 27 &  B  & 1.96  $\pm$ 0.26  &  \\
& BB116  & 2000-06-26 & 9 & 2007+404 & 2.40  $\pm$ 0.01  & 4.9  &  20 & 26 & DnC & 2.9   $\pm$ 2.3   &  \\
& AS786  & 2004-02-15 & 7 & 2015+371 & 2.76  $\pm$ 0.02  & 5.4  &  45 & 25 & CnB & 1.12  $\pm$ 0.18  &  \tablefootmark{c} \\
& BD110  & 2005-11-15 & 9 & 2007+404 & 2.29  $\pm$ 0.02  & 4.9  &  21 & 22 &  D  &         $<$ 2.1   &  S  \\
& AB1195 & 2006-01-26 & 8 & 2007+404 & 2.37  $\pm$ 0.05  & 4.9  &  31 & 16 & AD  & 0.58  $\pm$ 0.27  &  \\
& AB1210 & 2006-05-24 & 8 & 2007+404 & 2.21  $\pm$ 0.02  & 4.9  &  27 & 21 & BnA & 0.98  $\pm$ 0.05  &  \\
\multicolumn{5}{l}{{\bf 6 cm}}\\*
& CHUR   & 1980-05-23 & 8 & 2007+404 & 4.73  $\pm$ 0.03  &  4.9 &  47 & 22 & AD  & 1.33  $\pm$ 0.13  &   1x50 \tablefootmark{a}   \\
& CHUR   & 1980-05-24 & 8 & 2007+404 & 4.58  $\pm$ 0.03  &  4.9 &  47 & 22 & AD  & 1.18  $\pm$ 0.11  &   1x50 \tablefootmark{a} \\
& BIGN   & 1981-10-16 & 9 & 2007+404 & 4.16  $\pm$ 0.04  &  4.9 &  17 & 27 & C   & 0.8   $\pm$ 0.4   &   S,1x50   \\
& BECK   & 1982-03-27 & 9 & 2007+404 & 4.74  $\pm$ 0.01  &  4.9 & 228 & 27 & A   & 1.5   $\pm$ 0.4   &   S,1x50   \\
& BECK   & 1982-07-03 & 9 & 2007+404 & 4.95  $\pm$ 0.02  &  4.9 &   5 & 26 & A   &         $<$ 5.    &   S,1x50   \\
& BIEG   & 1982-08-26 & 9 & 2007+404 & 4.4   $\pm$ 0.1   &  4.9 &   9 & 24 & B   &         $<$ 3.    &   1x50 \\
& BECK   & 1982-12-21 & 9 & 2023+318 & 2.86  $\pm$ 0.01  &  9.6 &   9 & 24 & D   &         $<$ 2.4   &   S,1x50   \\
& AC42   & 1983-05-09 & 9 & 2007+404 & 4.46  $\pm$ 0.02  &  4.9 &  10 & 26 & C   &         $<$ 1.2   &   1x50   \\
& AC42   & 1983-08-22 & 9 & 2007+404 & 4.85  $\pm$ 0.01  &  4.9 &  14 & 25 & A   &         $<$ 4.    &   \\
& AB228  & 1983-08-27 & 9 & 2007+404 & 4.90  $\pm$ 0.03  &  4.9 &  18 & 27 & A   &         $<$ 3.    &   \\
& AB252  & 1983-10-30 & 9 & 2007+404 & 4.84  $\pm$ 0.02  &  4.9 &   9 & 26 & A   &         $<$ 4.    &   S \\
& AA28   & 1984-03-04 & 8 & 2007+404 & 4.27  $\pm$ 0.01  &  4.9 &  41 & 27 & CnB & 0.91  $\pm$ 0.10  &   \tablefootmark{a} \\
& AA28   & 1984-03-09 & 8 & 2007+404 & 4.33  $\pm$ 0.01  &  4.9 &  23 & 26 & CnB & 0.78  $\pm$ 0.13  &   \\
& AA29   & 1984-04-04 & 9 & 2007+404 & 4.21  $\pm$ 0.02  &  4.9 &  11 & 27 & C   &         $<$ 0.9   &   S \tablefootmark{a} \\
& VM59   & 1984-10-16 & 9 & 2202+422 & 2.72  $\pm$ 0.01  & 16.7 &  13 & 26 & D   &         $<$ 2.4   &   S \\
& AC116  & 1984-11-27 & 9 & 2007+404 & 4.50  $\pm$ 0.01  &  4.9 &  19 & 25 & A   &         $<$ 2.8   &   S \\
& AC116  & 1984-12-21 & 8 & 2007+404 & 4.63  $\pm$ 0.01  &  4.9 &   6 & 27 & A   & 0.56  $\pm$ 0.13  &   \tablefootmark{a} \\
& AF102  & 1985-05-09 & + & 2022+616 & 2.43  $\pm$ 0.01  & 20.4 &   9 & 26 & B   &         $<$ 1.3   &   S \\
& AA47   & 1985-06-12 & 7 & 2007+404 & 3.87  $\pm$ 0.01  &  4.9 &  24 & 22 & B   & 0.98  $\pm$ 0.27  &   \\
& AA47   & 1985-09-01 & 9 & 2007+404 & 3.53  $\pm$ 0.01  &  4.9 &   7 & 27 & C   &         $<$ 1.0   &   \\
& AA47   & 1985-09-19 & 9 & 2007+404 & 3.56  $\pm$ 0.01  &  4.9 &  11 & 25 & C   &         $<$ 0.8   &   \\
& VM115  & 1990-06-11 & 9 & 2202+422 & 3.24  $\pm$ 0.01  & 16.7 &  19 & 22 & A   &         $<$ 1.6   &   \\
& AG320  & 1991-03-06 & SE& 2007+404 & 2.73  $\pm$ 0.01  &  5.3 &  10 & 25 & D   & 1.4   $\pm$ 0.6   &   \\
& AB671  & 1993-01-21 & 9 & 2007+404 & 3.09  $\pm$ 0.01  &  4.9 &  19 & 24 & A   &         $<$ 2.3   &   S \\
& AB671  & 1993-01-24 & 9 & 2007+404 & 3.08  $\pm$ 0.03  &  4.9 &  18 & 27 & A   &         $<$ 2.8   &   S \\
& AB671  & 1993-01-29 & 9 & 2007+404 & 3.12  $\pm$ 0.02  &  4.9 &  19 & 27 & BnA & 2.1   $\pm$ 1.1   &   S \\
& AB671  & 1993-02-01 & 9 & 2007+404 & 3.11  $\pm$ 0.01  &  4.9 &   7 & 27 & BnA & 1.2   $\pm$ 0.7   &   S \\
& AB671  & 1993-02-14 & 9 & 2007+404 & 3.12  $\pm$ 0.02  &  4.9 &   8 & 27 & BnA & 2.2   $\pm$ 1.4   &      \\
& AS483  & 1993-05-01 & 98& 2007+404 & 3.04  $\pm$ 0.01  &  4.9 &  21 & 25 & B   & 1.33  $\pm$ 0.16  &   \tablefootmark{b} \\
& TST6CM & 1993-06-04 & 9 & 2052+365 & 5.2   $\pm$ 0.2   &  6.0 &  29 & 26 & CnB &         $<$ 0.8   &   X \\
& AR277  & 1994-04-17 & 9 & 2007+404 & 3.04  $\pm$ 0.01  &  4.9 &  23 & 15 & A   &         $<$ 4.    &   \\
& AR328  & 1995-04-27 & 9 & 2007+404 & 2.982 $\pm$ 0.003 &  4.9 &  11 & 16 & D   &         $<$ 2.3   &   \\
& AS644  & 1999-02-05 & 7 & 2007+404 & 2.6   $\pm$ 0.1   &  4.9 &  20 & 23 & DnC & 1.1   $\pm$ 0.5   &   \\
& AS644  & 1999-04-18 & 8 & 2015+371 & 1.728 $\pm$ 0.003 &  5.4 &   5 & 26 & D   & 0.93  $\pm$ 0.19  &   \\
& AW515  & 1999-06-08 & 9 & 2007+404 & 2.40  $\pm$ 0.03  &  4.9 &  10 & 21 & AD  &         $<$ 1.4   &   I,S \\
& AS786  & 2004-02-15 & 7 & 2015+371 & 2.778 $\pm$ 0.003 &  5.4 &  61 & 26 & CnB & 1.21  $\pm$ 0.28  &   \tablefootmark{c} \\
& AB1156 & 2005-02-04 & 8 & 2007+404 & 2.47  $\pm$ 0.01  &  4.9 &  31 & 25 & BnA & 1.28  $\pm$ 0.05  &   \\
& AB1156 & 2005-02-06 & 8 & 2007+404 & 2.50  $\pm$ 0.02  &  4.9 &  31 & 25 & BnA & 1.27  $\pm$ 0.06  &   \\
& AB1156 & 2005-02-08 & 8 & 2007+404 & 2.52  $\pm$ 0.01  &  4.9 &  31 & 25 & BnA & 1.11  $\pm$ 0.12  &   \\
& AB1156 & 2005-02-11 & 8 & 2007+404 & 2.52  $\pm$ 0.01  &  4.9 &  31 & 25 & BnA & 0.88  $\pm$ 0.07  &   \\
& AB1156 & 2005-02-13 & 8 & 2007+404 & 2.49  $\pm$ 0.02  &  4.9 &  31 & 24 & BnA & 0.76  $\pm$ 0.05  &   \\
& AB1156 & 2005-02-18 & 8 & 2007+404 & 2.44  $\pm$ 0.01  &  4.9 &  51 & 24 & B   & 0.64  $\pm$ 0.04  &   \\
& AB1156 & 2005-02-25 & 8 & 2007+404 & 2.49  $\pm$ 0.01  &  4.9 &  30 & 24 & B   & 1.11  $\pm$ 0.06  &   \\
& AB1156 & 2005-02-27 & 8 & 2007+404 & 2.49  $\pm$ 0.01  &  4.9 &  51 & 25 & B   & 0.98  $\pm$ 0.06  &   \\
& AB1156 & 2005-02-28 & 8 & 2007+404 & 2.47  $\pm$ 0.01  &  4.9 &  30 & 25 & B   & 1.29  $\pm$ 0.05  &   \\
& AB1156 & 2005-03-01 & 8 & 2007+404 & 2.410 $\pm$ 0.004 &  4.9 &  31 & 24 & B   & 1.14  $\pm$ 0.07  &   \\
& AB1156 & 2005-03-02 & 8 & 2007+404 & 2.44  $\pm$ 0.01  &  4.9 &  29 & 25 & B   & 1.26  $\pm$ 0.05  &   \\
& AB1156 & 2005-03-08 & 8 & 2007+404 & 2.44  $\pm$ 0.01  &  4.9 &  26 & 25 & B   & 0.72  $\pm$ 0.05  &   \\
& AB1156 & 2005-03-12 & 8 & 2007+404 & 2.43  $\pm$ 0.01  &  4.9 &  30 & 25 & B   & 0.73  $\pm$ 0.05  &   \\
& BD110  & 2005-11-15 & 9 & 2007+404 & 2.51  $\pm$ 0.01  &  4.9 &  18 & 21 & D   &         $<$ 1.9   &   S \\
\multicolumn{5}{l}{{\bf 20 cm}}\\*
& BIEG   & 1982-08-26 & 12& 2007+404 & 4.00  $\pm$ 0.04  &  4.8 & 53  & 24 & B   &         $<$ 3.1   &  \\
& AB228  & 1983-08-25 & 9 & 2007+404 & 4.19  $\pm$ 0.02  &  4.9 & 28  & 24 & A   &         $<$ 2.2   &  \\
& AB228  & 1983-08-27 & 12& 2007+404 & 4.31  $\pm$ 0.02  &  4.8 & 107 & 26 & A   &         $<$ 2.5   &  \\
& AA28   & 1984-03-09 & 8 & 2007+404 & 4.15  $\pm$ 0.02  &  4.9 & 26  & 26 & CnB & 1.14  $\pm$ 0.28  &  \\
& AC116  & 1984-11-27 & 9 & 2007+404 & 4.06  $\pm$ 0.02  &  4.9 & 33  & 25 & A   & 2.6   $\pm$ 0.8   &  \\
& AC116  & 1984-12-21 & 8 & 2007+404 & 3.98  $\pm$ 0.03  &  4.9 & 11  & 27 & A   & 1.29  $\pm$ 0.13  &  \tablefootmark{a} \\
& AT61   & 1985-04-04 & 9 & 2052+365 & 5.43  $\pm$ 0.01  &  6.0 & 14  & 25 & BnA &         $<$ 1.7   &  \\
& AA47   & 1985-09-19 & 9 & 2007+404 & 3.92  $\pm$ 0.04  &  4.9 & 13  & 26 & C   &         $<$ 4.    &  S \\
& AR170  & 1987-10-16 & 5 & 2052+365 & 5.20  $\pm$ 0.02  &  6.2 & 26  & 23 & BnA & 2.1   $\pm$ 1.2   &  \\
& AR170  & 1987-11-09 & 5 & 2052+365 & 5.07  $\pm$ 0.05  &  6.2 & 9   & 21 & BnA &         $<$ 4.    &  \\
& AM305  & 1990-07-05 & 80& 2007+404 & 2.84  $\pm$ 0.02  &  4.9 & 5   & 26 & BnA &         $<$ 2.8   &  2x25 \\
& AFTST  & 1991-08-11 & 21& 2052+365 & 5.15  $\pm$ 0.01  &  6.0 & 38  & 19 & A   &         $<$ 5.    &  \\
& AW311  & 1992-03-15 & 21& 2022+542 & 1.036 $\pm$ 0.002 & 13.2 & 29  & 27 & C   &         $<$ 5.    &  S \\
& TST6CM & 1993-06-04 & 9 & 2052+365 & 6.4   $\pm$ 0.3   &  6.0 & 13  & 27 & CnB &         $<$ 4.    &  S \\
& AL372  & 1996-03-28 & 5 & 2038+513 & 5.80              & 10.1 & 10  & 25 & C   &         $<$ 3.0   &  S,PH  \\
&        &            &   &          &                   &      &     &    &     &                   &  2x3.125  \\
& AA237  & 1999-02-01 & J2& 2052+365 & 5.31  $\pm$ 0.03  &  6.1 & 11  & 27 & C   &         $<$ 4.    &  \\
\hline
\end{longtable}
\tablefoot{
Column (1) gives the programme name, (2) the date of the observation,
(3) the source on which the observation was centred,
(4) the phase calibrator name (J2000 coordinates),
(5) the phase calibrator flux and 
(6) distance to the observation centre,
(7) the integration time on the source,
(8) the number of antennas that gave a usable signal,
(9) the configuration the VLA was in at the time of the observation,
(10) the measured flux and
(11) refers to the notes.
Many of the VLA observations were made in two sidebands, each of which has a 
bandwidth of 50 MHz; the exceptions are noted in column (11).
Upper limits are 3 $\times$ the RMS. \\
The details of column (3) are: the source on which the observation was centred
is mostly indicated by \citet{1956ApJ...124..530S} Cyg~OB2 numbers. 
Other indicators are:
`98' is in between Cyg~OB2 No.~8A and No.~9; `+' is listed in the observing log as
2031+411;
`80' as F80XF, `21' as 2032+41;
`J2' as 2033+4118\_2 and `SE' as TEVSE. \\
The notes in column (11) are: 
 `1x50' (or a similar code): bandwidth different from standard 2x50 MHz;
 `I':  stripes are visible on this image, complicating the flux measurement;
 `PH':  flux calibration on phase calibrator; 
 `S':  selfcalibration applied; 
 `X':    flux calibration is very uncertain; check on flux of phase calibrator suggests a factor of 2 uncertainty.
Column (11) also gives references for those
observations that have already been published in the literature:
\tablefoottext{a}{ \citet{1989ApJ...340..518B}} 
\tablefoottext{b}{ \citet{1998ApJS..118..217W}} 
\tablefoottext{c}{ \citet{2006A&A...454..625P}}.
}

\end{document}